\documentclass[preprint,groupedaddress,pre]{revtex4}
\begin{document}

\title{Two-degree-of-freedom Hamiltonian for\ the time-symmetric two-body
problem of the relativistic action-at-a-distance electrodynamics }
\author{Efrain Buksman Hollander and Jayme De Luca}
\email[corresponding author; email address:]{ deluca@df.ufscar.br}
\affiliation{Universidade Federal de S\~{a}o Carlos, \\
Departamento de F\'{\i}sica\\
Rodovia Washington Luis, km 235\\
Caixa Postal 676, S\~{a}o Carlos, S\~{a}o Paulo 13565-905}
\date{\today }

\begin{abstract}
We find a two-degree-of-freedom Hamiltonian for the time-symmetric problem
of straight line motion of two electrons in direct relativistic interaction.
This time-symmetric dynamical system appeared 100 years ago and it was
popularized in the 1940's by the work of Wheeler and Feynman in
electrodynamics, which was left incomplete due to the lack of a Hamiltonian
description. The form of our Hamiltonian is such that the action of a
Lorentz transformation is explicitly described by a canonical transformation
(with rescaling of the evolution parameter). The method is closed and
defines the Hamitonian in implicit form without power expansions. We outline
the method with an emphasis on the physics of this complex conservative
dynamical system. The Hamiltonian orbits are calculated numerically at low
energies using a self-consistent steepest-descent method (a stable numerical
method that chooses only the nonrunaway solution). The two-degree-of-freedom
Hamiltonian suggests a simple prescription for the canonical quantization of
the relativistic two-body problem.
\end{abstract}

\maketitle

\section{Introduction}

The class of equivariant dynamical systems under the Poincar\'{e} group has
enormous relevance to physics and yet, to date, only the one-body
relativistic motion is fully understood. Already with two bodies in
relativistic motion, one encounters the no-interaction theorem: a group
theoretical obstacle to the Hamiltonian description of relativistic
two-particle motion \cite{Currie}. The no-interaction theorem can be
overcome by covariant constraint dynamics \cite{Constraint}, but one is left
with the few cases where the constraint scheme closes. For example, for an
equivariant physical theory like the time-symmetric electrodynamics \cite%
{Fey-Whe}, a constraint description is unknown. In this paper we present a
reduction of the time-symmetric two-body problem of the relativistic
action-at-a-distance electrodynamics to a two-degree-of-freedom Hamiltonian
system along the nonrunaway solutions. The form of the Hamiltonian is such
that a Lorentz transformation is explicitly described by a canonical
transformation with rescaling of the evolution parameter. The Hamiltonian
orbits are calculated numerically by a numerically stable self-consistent
method that uses steepest-descent quenching and chooses only the nonrunaway
orbits.

In 1903, Schwarzchild proposed a relativistic type of interaction between
charges that was time reversible precisely because it involved retarded and
advanced interactions symmetrically \cite{Schwarz}. The same model
reappeared in the 1920s in the work of Tetrode and Fokker \cite%
{Tetrode-Fokker} and it finally became an interesting physical theory after
Wheeler and Feynman showed that this direct-interaction theory can describe
all the classical electromagnetic phenomena (i.e. the classical laws of
Coulomb, Faraday, Amp\`{e}re, and Biot-Savart) \cite{Fey-Whe,Leiter}.
Another accomplishment was that Wheeler and Feynman showed in 1945 that in a
certain limit where the electron practically interacts with a completely
absorbing universe, the response of this universe to the electron's field is
equivalent to the \emph{local} Lorentz-Dirac self-interaction theory \cite%
{Dirac} without the need of mass renormalization \cite{Fey-Whe}. It is
amusing to understand that the classical radiative phenomena of Maxwell's
electrodynamics can be described as a limiting case of \ this
direct-interaction theory (complete absorption is added to the theory as a
simple model to uncouple it from the detailed neutral-delay dynamics of the
other charges of the universe; for other limits see Ref. \cite{Narlikar}). \

For the relativistic two-body system of the action-at-a-distance
electrodynamics, general solutions are not known and the only known special
solution is the circular orbit for the attractive two-body problem, first
found in Ref. \cite{Schonberg} and later rediscovered in Ref. \cite{Schild}
( see also \cite{Hans} and \cite{Bonn}). Our problem has already been
studied: the symmetric motion of two electrons along a straight line [$%
-x_{2}(t)=x_{1}(t)\equiv x(t)$], which has the following equation in the
action-at-a-distance electrodynamics
\begin{equation}
m\frac{d}{dt}(\frac{v}{\sqrt{1-(v/c)^{2}}})=\frac{e^{2}}{2r^{2}}\left( \frac{%
1-v(t-r)/c}{1+v(t-r)/c}\right) +\frac{e^{2}}{2q^{2}}\left( \frac{1+v(t+q)/c}{%
1-v(t+q)/c}\right) ,  \label{eqmotion}
\end{equation}%
where $v(t)\equiv dx/dt$ is the velocity of the first electron, of mass $m$
and charge $e$, and $r$ and $q$ are the time-dependent delay and advance,
respectively, which are implicitly defined by the light-cone conditions%
\begin{eqnarray}
cr(t) &=&x(t)+x(t-r),  \label{lightcone} \\
cq(t) &=&x(t)+x(t+q),  \nonumber
\end{eqnarray}%
where $c$ is the speed of light. In general, a neutral-delay equation such
as Eqs. (\ref{eqmotion}) and (\ref{lightcone}) requires an initial function
as the initial condition, but for the special case of equations (\ref%
{eqmotion}) and (\ref{lightcone}) it was proved in 1979 that for
sufficiently low energies the Newtonian initial condition [ $x(0)=x_{o}$ and
$v(0)=v_{o}$ ] determines a unique symmetric solution that is globally
defined (i.e., that does not runaway at some point) \cite{Driver1,Driver2}.
This surprising uniqueness theorem reducing the initial condition from an
arbitrary function to two simple real numbers (initial position and
velocity) already suggests that the physical phase space could be isomorphic
to a two-degree-of-freedom Hamiltonian vector field, at least for low
velocities (which is what we find here). The first numerical method to solve
Eqs. (\ref{eqmotion}) and (\ref{lightcone}) was given in \cite{VonBaeyer}
and converged to solutions up to $v/c=0.94.$ Later another method \cite%
{Igor-pair}\ converged up to $v/c=0.99.$

In the following we present a method to find the nonrunaway solution of Eqs.
(\ref{eqmotion}) and (\ref{lightcone}) with a two-degree-of-freedom
Hamiltonian system. Our method is based on the physics; it starts from the
Fokker action and transforms the neutral-advance-delay equation into two
separate \emph{Hamiltonian} ordinary differential equations for the same
trajectory in two different foliations. The Hamiltonians are defined in
implicit form and can be solved explicitly in terms of the arbitrary ghost
functions by using the Hamilton-Jacobi theory. The condition that the two
solutions describe the same trajectory poses a functional problem involving
one of the ghost functions, with known asymptotic form, which must be solved
self-consistently and is the basis of our numerical calculation of the
Hamiltonian orbits. We outline the method with an emphasis on the physics
described by this complex conservative dynamical system. The paper is
organized as follows: in Sec. II we describe the bi-Lagrangian method and
solve explicitly for the motion resulting from the Hamiltonians. In Sec. III
we discuss some consequences of symmetry on the explicit solution of Sec. II
to reduce the number of arbitrary functions. In Sec. IV we determine an
equation to match the dynamics of one of the particles in the two
foliations, which turns out to involve only one of the ghost functions. In
Sec. V we use steepest-descent quenching to find the self-consistent
Hamiltonian orbits. In appendix 1 we prove the twice-monotonic property for
the arbitrary-mass case. Appendix 2 discusses an alternative covariant
derivation of the equal-mass case, and we also show here that the action of
a Lorentz transformation on the Hamiltonian is represented by a canonical
transformation with rescaling of the evolution parameter. In Sec. VI we give
the conclusions and discussion.

\section{Outline of the Method}

Here we consider the isolated two-body system with repulsive interaction,
away from the other charges of the universe, a conservative time-reversible
dynamical system in the action-at-a-distance electrodynamics. The equations
of motion for two bodies in the action-at-a-distance electrodynamics of
Wheeler and Feynman \cite{Fey-Whe}, henceforth called 1D-WF2B, our Eqs. (\ref%
{eqmotion}) and (\ref{lightcone}), are derived formally \cite{Staruskiewicz}
by extremizing the Schwarzschild-Tetrode-Fokker action functional
\begin{equation}
S_{F}=-\int m_{1}ds_{1}-\int m_{2}ds_{2}-e^{2}\int \int \delta
(||x_{1}-x_{2}||^{2})\dot{x}_{1}\cdot \dot{x}_{2}ds_{1}ds_{2},
\label{Fokker}
\end{equation}%
where $x_{i}(s_{i})$ represents the four-position of particle $i=1,2$
parametrized by its arc-length $s_{i}\,$, double bars stand for the
four-vector modulus $||x_{1}-x_{2}||^{2}\equiv (x_{1}-x_{2})\cdot
(x_{1}-x_{2})$, and the dot indicates the Minkowski scalar product of
four-vectors with the metric tensor $g_{\mu \nu }$\ ($%
g_{00}=1,g_{11}=g_{22}=g_{33}=-1$). The particles have masses $m_{1}$, $%
m_{2} $, common charge $e$, and in our units, $c=1$ \cite{Anderson}. The
formal conserved energy associated with the Poincar\'{e} invariance of the
Fokker Lagrangian (\ref{Fokker}) is discussed in Ref. \cite{Fey-Whe,Anderson}%
, a nonlocal expression involving an integral over a portion of the
trajectory, which is not useful to the present work, even though we start
from the same Lagrangian (\ref{Fokker}).

The starting point of our method is a transformation to new variables
\begin{eqnarray}
\xi _{1} &\equiv &t_{1}-x_{1},\qquad \zeta _{1}\equiv t_{1}+x_{1},
\label{defining} \\
\xi _{2} &\equiv &t_{2}-x_{2},\qquad \zeta _{2}\equiv t_{2}+x_{2}.  \nonumber
\end{eqnarray}%
As first noticed in Ref. \cite{Staruskiewicz}, this transformation splits
the action integral (\ref{Fokker}) into two separate local actions
\begin{equation}
S_{F}=\frac{1}{2}(S_{a}+S_{b}),  \label{splitFokk}
\end{equation}%
with

\begin{eqnarray}
S_{a} &=&-\int m_{1}(d\xi _{1}d\zeta _{1})^{1/2}-\int m_{2}(d\xi _{2}d\zeta
_{2})^{1/2}  \label{eqSa} \\
&&-e^{2}\int \int \frac{\delta (\zeta _{1}-\zeta _{2})}{|\xi _{1}-\xi _{2}|}%
(d\xi _{1}d\zeta _{2}+d\xi _{2}d\zeta _{1}),  \nonumber
\end{eqnarray}%
and
\begin{eqnarray}
S_{b} &=&-\int m_{1}(d\xi _{1}d\zeta _{1})^{1/2}-\int m_{2}(d\xi _{2}d\zeta
_{2})^{1/2}  \label{eqSb} \\
&&-e^{2}\int \int \frac{\delta (\xi _{1}-\xi _{2})}{|\zeta _{1}-\zeta _{2}|}%
(d\xi _{1}d\zeta _{2}+d\xi _{2}d\zeta _{1}).  \nonumber
\end{eqnarray}%
It should be noticed that the double integral of Eq. (\ref{Fokker}) is
reduced, after integration of the $\delta $ function, to a single integral
over the parameter of particle $1$, with particle $2$ contributing only at
the advanced and retarded positions, this being precisely the reason for the
nonlocality of the theory, as illustrated in Eq. (\ref{eqmotion}). The
usefulness of parametrization (\ref{defining}) is that it naturally breaks
the double integral of (\ref{Fokker}) into two integrals, each involving a
different $\delta $ function, and integration over each $\delta $ function
couples particle $1$ with particle $2$ at \emph{either }the advanced
position [the double integral included in $S_{a}$ of Eq. (\ref{eqSa})] \emph{%
or} at the retarded position [the double integral included in $S_{b}$ of \
Eq. (\ref{eqSb})]. For example, in action $S_{a}$ of Eq. (\ref{eqSa}), the
nonzero contribution of the $\delta $ function occurs where the parameters $%
\zeta _{1}$ and $\zeta _{2}$ take equal values $\zeta _{1}=\zeta _{2}$ =$%
\zeta $, and this $\zeta $ is the natural independent parameter of the local
action $S_{a}$ (this parametrization is often named front form of dynamics
in the literature and we henceforth call it type $a$ foliation). \ For
action $S_{b}$ of Eq. (\ref{eqSb}), integration over the $\delta $ function
produces a contribution to the integral only where the two parameters $\xi
_{1}$ and $\xi _{2}$ are equal, and $\xi _{1}=\xi _{2}=\xi $ is the natural
time parameter of action $S_{b}$ (henceforth called type $b$ foliation). To
gain some insights into the two types of foliation, we notice that with type
$a$, the particles are automatically in the light-cone condition $%
(x_{1}-x_{2})^{2}-(t_{1}-t_{2})^{2}=0$, \ particle $2$ always being ahead of
particle $1$ in time after the choice $x_{1}-x_{2}>0$, with the light-cone
distance being
\begin{equation}
r_{a}=-\frac{1}{2}\bigskip (\xi _{1}-\xi _{2}).  \label{definera}
\end{equation}%
With type $b$ parametrization the particles are also in the light-cone
condition, with particle $2$ behind in time and the light-cone distance
being
\begin{equation}
r_{b}=\frac{1}{2}\bigskip (\zeta _{1}-\zeta _{2}).  \label{definerb}
\end{equation}

The first heuristic guide for this work, as first noticed in Ref. \cite%
{Staruskiewicz}, is the simplicity of the Euler-Lagrange problem for partial
action (\ref{eqSa}) : after expressing action (\ref{eqSa}) in terms of the
time-like parameter $\zeta $, it is easily verified that the associated
Euler-Lagrange equation is a simple ordinary differential equation (not a
delay equation anymore!). The Euler-Lagrange problem for action (\ref{eqSb})
is analogous, with $\zeta $ replaced by $\xi .$ To avoid confusion, we
henceforth define that a Lagrangian has a \emph{local} form when the
associated Euler-Lagrange problem is defined by an ordinary differential
equation. In searching for a local Lagrangian problem, we could try to
extremize each of the partial action functionals of Eqs. (\ref{eqSa}) and (%
\ref{eqSb}) and obtain a trajectory by solving the Euler-Lagrange equation
for either $\delta S_{a}=0$ or $\delta S_{b}=0.$ Each separate minimization,
in general, yields a different trajectory, which is the time-asymmetric
problem studied in several works \cite{Rudd}. The main idea of our method is
that \emph{if }these two trajectories turn out to be equal, this common
trajectory also extremizes the original action integral (\ref{Fokker}), as $%
\delta S_{F}=\frac{1}{2}\delta S_{a}+\frac{1}{2}\delta S_{b}=0+0=0.$ Simply
formulated as above, the problem turns out to be impossible; and it is
possible to prove that the two separate solutions can never describe the
same orbit. To overcome this difficulty we need to postulate a more general
bi-Lagrangian problem by simultaneously solving%
\begin{equation}
\delta S_{a}=\delta G,  \label{var1}
\end{equation}%
and
\begin{equation}
\delta S_{b}=-\delta G,  \label{var2}
\end{equation}%
with $G$ being a so far undetermined Lagrangian. A trajectory that satisfies
Eqs. ($\ref{var1}$) and ($\ref{var2}$) will also extremize the Fokker action
(\ref{Fokker}), a simple consequence of Eqs. (\ref{splitFokk}), (\ref{var1})
and (\ref{var2}):%
\begin{equation}
\delta S_{F}=\frac{1}{2}\delta S_{a}+\frac{1}{2}\delta S_{b}=\frac{1}{2}%
\delta G-\frac{1}{2}\delta G=0.  \label{consequence}
\end{equation}

Our first task is to find a sufficiently general Lagrangian $G$ \ such that
Eqs. ($\ref{var1}$) and ($\ref{var2}$) yield the same trajectory. Once we
are trying to avoid delay equations, it is desirable that the Euler-Lagrange
Eqs. for (\ref{var1}) and (\ref{var2}) be ordinary differential equations,
which is the heuristic guide for choosing the functional $G$. A functional $%
G $ that leaves the two separate problems (\ref{var1}) and (\ref{var2}) in
the local form is henceforth called a bilocal ghost Lagrangian $G$. Here we
consider symmetric and time-reversible solutions of Eq. (\ref{eqmotion})
only, but for the variational calculus that follows, it is necessary to
study such an orbit immersed in a family of orbits, defined as follows: A
time-reversible orbit naturally defines a preferred frame; the Lorentz frame
where the orbit is time-reversible, and we henceforth call it the center of
mass frame (CMF). We consider in the CMF the family of all orbits such that
the trajectories of electrons $1$ and $2$ are both time-reversible but not
necessarily equal (non-symmetric orbits) [$x_{1}(-t)=x_{1}(t)$] and [$%
x_{2}(-t)=x_{2}(t)$ ], and with the physical property that both the advanced
and retarded distances decrease monotonically to a point of minimum and then
start increasing monotonically again, as illustrated in Fig. 1. We
henceforth call this family of orbits the CMF family. The fact that the
solution of \ Eqs. (\ref{eqmotion}) and (\ref{lightcone}) has this piecewise
monotonic property is a consequence of the velocity being a monotonic
function of time, which was proved in Ref. \cite{Driver1}\ for sufficiently
low velocity orbits (in Appendix 1 we prove this assertion for the
arbitrary-mass case). We henceforth refer to a CMF orbit as a
twice-monotonic orbit. Since the solution we are looking for is symmetric
and time-reversible, it obviously belongs to the CMF family, and since this
solution extremizes Eq. (\ref{Fokker}) in the family of all orbits, it
obviously does so restricted to the CMF family. A symmetric and
time-reversible orbit seen in a Lorentz frame other than the CMF has the
property that the future of electron $1$ is the past of electron $2$ and
vice-versa. In the following we restrict the analysis of the different-mass
case to the CMF. For a general covariant derivation of the equal-mass case
see Appendix 2.

In the following we prove four integral identities for the orbits of the CMF
family, which are later used to construct the bilocal ghost Lagrangian. The
action of the time-reversal operation on orbits of the CMF family can be
shown to be the following map: $\zeta _{1,2}\rightarrow $ $-\xi _{1,2}$ , $%
\xi _{1,2}\rightarrow -\zeta _{1,2}$, $r_{a}\rightarrow r_{b}$ and it is
worth noticing that time reversal maps type a parametrization onto type $b$
and vice-versa. In this work we ignore questions of convergence and define
all integrals formally from $-\infty $ to $\infty ,$ an ambiguity inherited
from the Wheeler-Feynman theory and discussed in Ref. \cite{Staruskiewicz}.
The simplest type of integral identity we shall use, valid for an arbitrary
function $\phi (x)$ of the real variable, is%
\begin{equation}
\int_{a}\phi (\zeta )d\zeta =\int_{b}\phi (\xi )d\xi .  \label{defphi}
\end{equation}%
The lower index of the integral denotes the parametrization type, and the
above identity is trivial, as with either type a or type $b$ the parameter
runs from $-\infty $ to $\infty $ ($\zeta $ for type $a$ and $\xi $ for type
$b$). It is also interesting to look at Eq. (\ref{defphi}) as a consequence
of the coordinate transformation induced by the time-reversal symmetry of
the CMF family ( $\zeta \rightarrow -\xi )$. In the same way, we can prove
in the CMF the following integral identity, involving an arbitrary function $%
V(x)$ of the real variable

\begin{equation}
\int_{a}V(\zeta )(\frac{d\xi _{1}}{d\zeta }+\frac{d\xi _{2}}{d\zeta })d\zeta
=\!\int_{b}V(\xi )(\frac{d\zeta _{1}}{d\xi }+\frac{d\zeta _{2}}{d\xi })d\xi ,
\label{defV}
\end{equation}%
The combination $(d\xi _{1}d\zeta _{2}+d\xi _{2}d\zeta _{1})$ is the time
reversible Lorentz-invariant area element that appeared naturally in Eqs. (%
\ref{eqSa}) and (\ref{eqSb}).\ Last, the same time-reversal action ($\zeta
_{1,2}\rightarrow -\xi _{1,2},\xi _{1,2}\rightarrow -\zeta _{1,2}$) on the
CMF family produces the following identities for arbitrary functions $\alpha
(\zeta )$ and $\beta (\zeta )$ of the real variable;%
\begin{eqnarray}
\int_{a}\alpha _{d,t}(\zeta )(d\xi _{1}d\zeta _{1})^{1/2} &=&\int_{b}\alpha
_{d,t}(\xi )(d\xi _{1}d\zeta _{1})^{1/2},  \label{defab} \\
\int_{a}\beta _{d,t}(\zeta )(d\xi _{2}d\zeta _{2})^{1/2} &=&\int_{b}\beta
_{d,t}(\xi )(d\xi _{2}d\zeta _{2})^{1/2}.  \nonumber
\end{eqnarray}%
The above identities suggest that we use a ghost Lagrangian $G$ of type%
\begin{equation}
G=\int_{a}[\phi (\zeta )+\frac{1}{2}V(\zeta )(\dot{\xi}_{1}+\dot{\xi}%
_{2})+\alpha (\zeta )\sqrt{\dot{\xi}_{1}}+\beta (\zeta )\sqrt{\dot{\xi}_{2}}%
]d\zeta .  \label{generalGa}
\end{equation}%
Notice that the dot over $\xi _{1,2}$ in equation (\ref{generalGa})
indicates the derivative respect to $\zeta $ ( the time-parameter of case $a$%
). This $G$ is in the local form when added to $S_{a\text{ }}$, where $\zeta
$ plays the role of the time parameter and the coordinates are $\xi _{1}$, $%
\xi _{2}$. When this same $G$ is subtracted from action $S_{b}$, the
integral identities allow us to express $G$ as
\begin{equation}
G=\int_{b}[\phi (\xi )+\frac{1}{2}V(\xi )(\dot{\zeta}_{1}+\dot{\zeta}%
_{2})+\alpha (\xi )\sqrt{\dot{\zeta}_{1}}+\beta (\xi )\sqrt{\dot{\zeta}_{2}}%
]d\xi ,  \label{generalGb}
\end{equation}%
which is also in the local form for action $S_{b}$, with $\xi $ being the
time parameter and the coordinates being $\zeta _{1}$ and $\zeta _{2}.$
Notice that the dot over $\zeta _{1,2}$ in equation (\ref{generalGb})
indicates derivative respect to $\xi $ (the time-parameter of case $b$). One
could in principle add more general parametrization-invariant terms to $G$ ;
for example, terms involving the integration element $(d\xi _{1}d\zeta
_{1})^{1/4}(d\xi _{2}d\zeta _{2})^{1/4}$ or any highly composite term, and
the inversion to the Hamiltonian formalism would involve several branches.
Lagrangian (\ref{generalGa}) is the most general ghost Lagrangian whose
associated Hamiltonian involves quadratic rational functions of the momenta,
and should suffice if the orbit has only two monotonic branches,
corresponding to the two elements of the Galois group of a quadratic
equation. The need for only four functions becomes also clearer later on,
when we find that there are four determining equations involving these four
arbitrary functions. We notice also that $\phi $ is defined up to a constant
in Eqs. (\ref{generalGa}) and (\ref{generalGb}), which is also true of $V$,
as adding a constant to $V$ simply adds a total time derivative to $G$ (a
gauge transformation). There is also no gain in generality if one defines a
general linear term like $(V_{1}\dot{\zeta}_{1}+V_{2}$ $\dot{\zeta}_{2})$ in
Eq. (\ref{generalGb}), as this is also a trivial transformation of the case
we used.

In the following we guide the reader to a division of the phase space into
two disjoint regions, as our constructive method defines one Hamiltonian for
each separate region as an implicit function of phase space: The condition $%
\dot{r}=0$ divides the phase space of a twice-monotonic orbit in two
separate regions according to whether $\dot{r}>0$ or $\dot{r}<0$ (in
Appendix 2 we show that this splitting is actually a covariant splitting for
the equal-mass case). The change from $\zeta $ to $r_{a}$ is one to one in
each of the two regions of a twice-monotonic orbit $\zeta \in \lbrack
-\infty ,-|\zeta _{c}|]$ and $\zeta \in \lbrack -|\zeta _{c}|,\infty ],$ as
can be seen from Figure 1, and this naturally splits all integrals into two.
For example, the left integral of identity (\ref{defphi}) splits as follows:

\begin{equation}
\int_{a}\phi (\zeta )d\zeta =\int_{-\infty }^{-|\zeta _{c}|}\phi
_{d}(r_{a})d\zeta +\int_{-|\zeta _{c}|}^{\infty }\phi _{t}(r_{a})d\zeta .
\label{intSa}
\end{equation}%
The above integral identity involves two arbitrary functions, and the
subscript $t$ (as in turn) indicates that function $\phi _{t}$ is defined in
the region of phase space where the trajectory of particle $1$ includes a
turning point (see Fig. 1), while subscript $d$ (as in direct) \ indicates
that $\phi _{d}$ is defined in the region of phase space where the
trajectory of particle $1$ is without a turning point (see Fig. 1). The same
integral can be expressed for type $b$ applying the time-reversal change of
variable $\zeta \rightarrow -\xi $ \ to the right side of Eq. (\ref{intSa}),
which maps $r_{a}$ to $r_{b}$ and maps the critical point $\zeta =-|\zeta
_{c}|$ of the $\zeta $ parametrization to the critical point $\xi =|\zeta
_{c}|$ of the $\xi $ parametrization,

\begin{equation}
\int_{b}\phi (\xi )d\xi =\int_{-\infty }^{|\zeta _{c}|}\phi _{t}(r_{b})d\xi
+\int_{-|\zeta _{c}|}^{\infty }\phi _{d}(r_{b})d\xi .  \label{intSb}
\end{equation}%
These two portions are indicated in Fig. 1 for both case $a$ and case $b.$
In the following we split all integrals of the ghost Lagrangians (\ref%
{generalGa}) and (\ref{generalGb}) in two, which we indicate with subscripts
$t$ and $d$ in the same way as of Eqs. (\ref{intSa}) and (\ref{intSb}). The
usefulness of the above splitting of the phase space is that one can express
all functions in the ghost Lagrangian as functions of the light-cone
distance (\ref{definera}) in each region, such that the Lagrangian becomes
independent of the time-like parameter and allows the existence of a
conserved energy $E$.

Our next task is to solve each separate problem for each separate Lagrangian
(and for each region of the phase space). After inclusion of the ghost
Lagrangian, the problem $\delta S_{a}=\delta G$ \ in the CMF family implies
the Euler-Lagrange equations for $L_{a}=S_{a}-G$,

\begin{equation}
L_{a}^{t,d}=-\int [M_{1a}^{t,d}\sqrt{\dot{\xi}_{1}}+M_{2a}^{t,d}\sqrt{\dot{%
\xi}_{2}}+(\frac{e^{2}}{|\xi _{1}-\xi _{2}|}+\frac{1}{2}V_{t,d}(r_{a}))(\dot{%
\xi}_{1}+\dot{\xi}_{2})+\phi _{t,d}(r_{a})]d\zeta ,  \label{Lagrangian-a}
\end{equation}%
where $M_{1a}^{t,d}\equiv m_{1}+\alpha _{t,d}(r_{a})$ and $%
M_{2a}^{t,d}\equiv m_{2}+\beta _{t,d}(r_{a})$ and the Lagrangian can be
uniquely inverted in each branch to produce a Hamiltonian, because of the
monotonic property.\ The problem $\delta S_{b}=-\delta G$ is described by $%
L_{b}=S_{b}+G$,

\begin{equation}
L_{b}^{t,d}=-\int [M_{1b}^{t,d}\sqrt{\dot{\zeta}_{1}}+M_{2b}^{t,d}\sqrt{\dot{%
\zeta}_{2}}+(\frac{e^{2}}{|\zeta _{1}-\zeta _{2}|}-\frac{1}{2}%
V_{t,d}(r_{b}))(\dot{\zeta}_{1}+\dot{\zeta}_{2})-\phi _{t,d}(r_{b})]d\xi ,
\label{Lagrangian-b}
\end{equation}%
with $M_{1b}^{t,d}\equiv m_{1}-\alpha _{t,d}(r_{b})$ and $M_{2b}^{t,d}\equiv
m_{2}-\beta _{t,d}(r_{b})$ . We have introduced eight arbitrary ghost
functions: $\phi _{t,d}\,,$ $V_{t,d}\,,\alpha _{t,d}$, and $\beta _{t,d}\,,$
four for each separate region of phase space, and we notice that these ghost
functions enter with a plus sign in case $a$ and with a minus sign in case $%
b.$ The Hamiltonian in each case is given by
\begin{equation}
H_{a}=\frac{-1}{4}\{\frac{M_{1a}^{2}}{(p_{1}+\frac{1}{2}V+\frac{e^{2}}{|\xi
_{1}-\xi _{2}|})}+\frac{M_{2a}^{2}}{(p_{2}+\frac{1}{2}V+\frac{e^{2}}{|\xi
_{1}-\xi _{2}|})}\}-\phi (r_{a}),  \label{hamia}
\end{equation}%
and
\begin{equation}
H_{b}=\frac{-1}{4}\{\frac{M_{1b}^{2}}{(p_{1}-\frac{1}{2}V+\frac{e^{2}}{%
|\zeta _{1}-\zeta _{2}|})}+\frac{M_{2b}^{2}}{(p_{2}-\frac{1}{2}V+\frac{e^{2}%
}{|\zeta _{1}-\zeta _{2}|})}\}+\phi (r_{b}).  \label{hamib}
\end{equation}%
We have omitted the subscripts but it should be kept in mind that each of
the above Hamiltonians is defined separately in each region of the phase
space, a separation that will be useful when we come to the symmetry
considerations. Notice that the Hamiltonian $H_{a}$ depends only on $r_{a}=-%
\frac{1}{2}(\xi _{1}-\xi _{2})$, which implies that $P_{a}=p_{1}+p_{2}$ is a
constant of motion. For type $b$ parametrization, Hamiltonian $H_{b}$
depends only on $r_{b}=\frac{1}{2}(\zeta _{1}-\zeta _{2})$, implying the
constant $P_{b}=p_{1}+p_{2}$. The constant $P_{a}=p_{1}+p_{2}$ suggests a
canonical change of variables for Hamiltonian $H_{a}$ , defined by
\begin{eqnarray}
X &\equiv &\frac{1}{2}(\xi _{1}+\xi _{2}),\quad P\equiv p_{1}+p_{2}
\label{canonical} \\
x &\equiv &\frac{1}{2}(\xi _{1}-\xi _{2}),\quad p=p_{1}-p_{2}.  \nonumber
\end{eqnarray}%
For type $b$ we use the analogous transformation with $\xi $ replaced by $%
\zeta $ in the above formulas. One can use Eq. (\ref{canonical}) to express $%
p_{1}$ and $p_{2}$ of Eq. $(\ref{hamia})$ in terms of the constant $P=P_{a}$
and the relative momentum $p$, and substitution into the condition $%
H_{a}=E_{a}$ yields a quadratic equation for $p$, with solutions%
\begin{equation}
p_{a}=\frac{\Delta _{a}}{(E_{a}+\phi )}\pm \sqrt{\left( P_{a}+\frac{Q_{a}}{%
(E_{a}+\phi )}+V(r)+\frac{e^{2}}{r_{a}}\right) ^{2}+\left( \frac{\Delta
_{a}^{2}-Q_{a}^{2}}{(E_{a}+\phi )^{2}}\right) },  \label{Coulomb}
\end{equation}%
where $Q_{a}\equiv \frac{1}{4}(M_{1a}^{2}+M_{2a}^{2})$ and $\Delta
_{a}\equiv \frac{1}{4}(M_{2a}^{2}-M_{1a}^{2})$ and $r_{a}=|x|.$ The
separation for case $b$ is analogous.

So far we have shown that any common solution of Hamiltonians (\ref{hamia})
and (\ref{hamib}) is also a solution of the original advance-delay problem
of Eqs. (\ref{eqmotion}) and (\ref{lightcone}), for arbitrarily given
potentials $\phi _{t,d},V_{t,d}\,,\alpha _{t,d}$, and $\beta _{t,d}$. It
turns out that, even if we guessed the four potentials correctly,
Hamiltonians (\ref{hamia}) and (\ref{hamib}) would have only a single
trajectory in common for each given set of potentials (this becomes clear in
the numerical work of Sec.V). This obstacle can be overcome with the
Hamiltonian formalism if we generalize the potentials of Eqs. (\ref{hamia})
and (\ref{hamib}) to implicit functions of the energy $E_{a}=H_{a}$ in case $%
a$ and of $E_{b}=H_{b}$ in case $b.$ For example, the potential $\phi $ is
generalized to $\phi =\phi (r_{a},E_{a})$ in case $a$ and to $\phi =\phi
(r_{b},E_{b})$ in case $b$ (an analogous generalization goes for $V$, $%
\alpha $, and $\beta $). This generalized Hamiltonian is still a function of
phase space, because $E$ itself is a function of phase space, even though it
is now only implicitly defined by Eqs. (\ref{hamia}) and (\ref{hamib}). In
this generalization, for each given orbit, of energy $E_{o}$, we still
define the ghost Lagrangians with (\ref{Lagrangian-a}) and (\ref%
{Lagrangian-b}) using \emph{fixed form }potentials: $\phi =\phi (r,E_{o})$; $%
V=V(r,E_{o})$; $\alpha =\alpha (r,E_{o})$; and $\beta =\beta (r,E_{o})$. By
construction, these generalized ghost Lagrangians have only a single orbit
in common with the generalized Hamiltonians, but it is essential that such
provisional Lagrangians exist, such that we can prove that the Hamiltonian
equations associated to (\ref{hamia}) and (\ref{hamib}) reduce to Eq. (\ref%
{eqmotion}), which is accomplished by using Eq. (\ref{consequence}) with
fixed form potentials. After that we can dispose of the Lagrangians. On the
Hamiltonian side, if we are changing the potentials with $E$ the Hamiltonian
equations of motion derived from (\ref{hamia}) and (\ref{hamib}) pick an
extra term proportional to the derivative of the Hamiltonian with respect to
$E$ (due to the implicit dependence). We must therefore supplement a
condition that this derivative vanishes along the orbit in each case, which
in case $a$ reads
\begin{equation}
\frac{\partial H_{a}(p,P,r,E_{a})}{\partial E_{a}}=0,  \label{firstwo-1}
\end{equation}%
and, in case $b$
\begin{equation}
\frac{\partial H_{b}(p,P,r,E_{b})}{\partial E_{b}}=0.  \label{firstwo-2}
\end{equation}%
In Eq. (\ref{firstwo-1}), the derivative $\partial H_{a}/\partial E_{a}=0$
should hold on the energy shell $H_{a}=E_{a}$ and in Eq. (\ref{firstwo-2})
the derivative $\partial H_{b}/\partial E_{b}=0$ should hold on the energy
shell $H_{b}=E_{b}.$ Elimination of the relative momentum $p_{a}$ from $%
H_{a}=E$ yields Eq. (\ref{Coulomb}) and substitution of (\ref{Coulomb}) into
(\ref{firstwo-1}) yields a partial differential equation (PDE) involving the
four potentials. An analogous PDE results for (\ref{firstwo-2}) in case $b$
such that Eqs. (\ref{firstwo-1}) and (\ref{firstwo-2}) define two partial
differential equations involving the four arbitrary potentials in each
region (variables of the partial differential equations are $r$, $P$, and $E$%
). Rigorously, the generalization to implicitly defined Hamiltonians
proceeds only if the time-reversal operation also maps $E_{a}$ into $E_{b}$.
For that we notice that $\phi $ enters with a plus sign in $H_{a}$ [equation
(\ref{hamia})] and with a minus sign in $H_{b}$ [equation (\ref{hamib})],
and the required symmetry can be accomplished by adding an energy dependent
constant to $\phi $. We conclude this paragraph stressing that the
generalized ghost Lagrangians were only a provisional artifact\emph{\ en
route} to the eventual derivation of the Hamiltonians\ (\ref{hamia}) and (%
\ref{hamib}) from a variational argument with use of symmetry. It should be
clear that after we generalize Eqs. (\ref{hamia}) and (\ref{hamib}) to
implicit dependence and postulate Eqs. (\ref{firstwo-1}) and (\ref{firstwo-2}%
) we can no longer go back to the simple provisional ghost Lagrangians, and
our constructive approach is \emph{essentially} left with an
implicitly-defined bi-Hamiltonian system.

In the following we show that even with Hamiltonian (\ref{hamia}) defined in
the implicit form we can write out the motion explicitly: This explicit
solution is accomplished in the manner of Hamilton-Jacobi, by use of a
canonical transformation with a generating function $S$ given by
\begin{equation}
S=PX+W(x,P,E)-E\zeta ,  \label{Jacobi}
\end{equation}%
where the function $W(x,P,E)$ is defined by integration from the condition $%
p=\partial W/\partial x$ , with $p$ given by Eq. (\ref{Coulomb}). This
canonical transformation is defined such that the new momentum associated
with the old variable $X$ is the same old constant $P$ $=\partial S/\partial
X$ and the other new momentum is the energy $E$ (with this last definition
we exploit the fact that $E$ is already one argument of the potentials). We
choose $S$ in the manner of Hamilton-Jacobi such that the new Hamiltonian
vanishes: $\ K=H+\frac{\partial S}{\partial \zeta }=0.$ As the Hamiltonian
is zero, the new coordinates are defined simply by two constants $X_{0\text{
}}$and $C_{0}$

\begin{eqnarray}
X_{0} &=&\partial S/\partial P=X+\partial W/\partial P  \label{Hami-Jaco} \\
\quad C_{0} &=&-\partial S/\partial E=\zeta -\partial W/\partial E  \nonumber
\end{eqnarray}%
The above equations for type $a$ define $\zeta $ and $X$ as functions of the
variable $r_{a}\equiv |x|$, and provide the complete solution of the
Hamiltonian motion. For further use, it is interesting to take the
differentials of Eq. (\ref{Hami-Jaco}) relative to $x,$%
\begin{eqnarray}
dX &=&-(\partial ^{2}W/\partial x\partial P)dx=-(\partial p/\partial P)dx,
\label{contact} \\
d\zeta &=&(\partial W/\partial x\partial E)dx=(\partial p/\partial E)dx,
\nonumber
\end{eqnarray}%
where we have used $p=\partial W/\partial x$ (definition of the
Hamilton-Jacobi transformation) and exchanged the partial derivatives. The
explicit form of the differential for the trajectory is obtained using Eq. (%
\ref{defining}) to relate particle coordinates to $X$ and $\zeta $ and using
(\ref{contact}) to relate $dX$ and $d\zeta $ to $dx.$ For type $a$
parametrization the explicit solution is
\begin{eqnarray}
dt_{1a} &=&\frac{1}{2}(d\zeta _{a}+dX_{a}+dx_{a})=\frac{1}{2}(\frac{\partial
p_{a}}{\partial P}-\frac{\partial p_{a}}{\partial E}-1)dr_{a},
\label{tablea} \\
dt_{2a} &=&\frac{1}{2}(d\zeta _{a}+dX_{a}-dx_{a})=\frac{1}{2}(\frac{\partial
p_{a}}{\partial P}-\frac{\partial p_{a}}{\partial E}+1)dr_{a},  \nonumber \\
dx_{1a} &=&\frac{1}{2}(d\zeta _{a}-dX_{a}-dx_{a})=\frac{1}{2}(1-\frac{%
\partial p_{a}}{\partial P}-\frac{\partial p_{a}}{\partial E})dr_{a},
\nonumber \\
dx_{2a} &=&\frac{1}{2}(d\zeta _{a}-dX_{a}+dx_{a})=-\frac{1}{2}(\frac{%
\partial p_{a}}{\partial P}+\frac{\partial p_{a}}{\partial E}+1)dr_{a},
\nonumber
\end{eqnarray}%
where we have also used $dx_{a}=-dr_{a}$. Analogously for type $b$ $%
(dx_{b}=dr_{b})$ we obtain the explicit solution%
\begin{eqnarray}
dt_{1b} &=&\frac{1}{2}(d\xi _{b}+dX_{b}+dx_{b})=\frac{1}{2}(\frac{\partial
p_{b}}{\partial E}-\frac{\partial p_{b}}{\partial P}+1)dr_{b},
\label{tableb} \\
dt_{2b} &=&\frac{1}{2}(d\xi _{b}+dX_{b}-dx_{b})=\frac{1}{2}(\frac{\partial
p_{b}}{\partial E}-\frac{\partial p_{b}}{\partial P}-1)dr_{b},  \nonumber \\
dx_{1b} &=&\frac{1}{2}(-d\xi _{b}+dX_{b}+dx_{b})=\frac{1}{2}(1-\frac{%
\partial p_{b}}{\partial P}-\frac{\partial p_{b}}{\partial E})dr_{b},
\nonumber \\
dx_{2b} &=&\frac{1}{2}(-d\xi _{b}+dX_{b}-dx_{b})=-\frac{1}{2}(\frac{\partial
p_{b}}{\partial P}+\frac{\partial p_{b}}{\partial E}+1)dr_{b}.  \nonumber
\end{eqnarray}%
We recall that Eqs. (\ref{tablea}) and (\ref{tableb}) give the explicit
solution in terms of $p(r,E,P)$ as given by Eq. (\ref{Coulomb}).

\bigskip

\section{\protect\bigskip\ Symmetry conditions for the equal-mass case}

In this section we discuss only the equal-mass case, and for that we set $%
m_{1}=m_{2}=m=1$ and allow only the charge to be arbitrary, from which the
general $m$ and $c$ case can be recovered by simply replacing $e^{2}$ by $%
(e^{2}/mc^{2})$. We henceforth set $e=1$ as well, which can be accomplished
by a rescaling of distances accompanied by a rescaling of time to keep $c=1$
. In the following we derive general symmetry relations involving the eight
arbitrary functions $\alpha _{t,d}\,,\beta _{t,d}\,,V_{t,d}$ , and $\phi
_{t,d}$ . Formula (\ref{Coulomb}) for $p_{a}$ is the solution of a quadratic
equation and defines two different functions $p_{a}$ by taking the plus and
minus signs of the square root. It is easy to show that at the branch point
the square root vanishes, so that a single branch of the square root
describes each of the $t$ and $d$ physical regions of phase space as
indicated in Fig. 1, henceforth indicated by $p_{a}^{t}$ and $p_{a}^{d}$ (on
$p_{a,b}$ we use superscripts to indicate branch type, to avoid overloaded
notation, but with the ghost functions we keep using subscripts). We have
assumed that the orbit is time-reversible in the CMF and, to be consistent
with that, time reversal must map each branch of the type $a$ trajectory of
particle $1$ onto a branch of its type $b$ trajectory, with the
corresponding velocities transforming like $%
v_{1a,b}^{d,t}(r)=-v_{1b,a}^{d,t}(r)$ for $r\in \lbrack r_{o},\infty ],$ as
illustrated in Fig. 2. In an analogous way, for particle $2$ we should have $%
v_{2a,b}^{d,t}(r)=-v_{2b,a}^{d,t}(r)$ for $r\in \lbrack r_{o},\infty ].$
These two symmetry conditions, when expressed in terms of $\partial
p/\partial P$ and $\partial p/\partial E$ using Eqs. (\ref{tablea}) and (\ref%
{tableb}), imply the four conditions

\begin{eqnarray}
\frac{\partial p_{a,b}^{t,d}}{\partial P} &=&\frac{\partial p_{b,a}^{t,d}}{%
\partial P},  \label{condi2-1} \\
\frac{\partial p_{a,b}^{t,d}}{\partial E} &=&\frac{\partial p_{b,a}^{t,d}}{%
\partial E}.  \label{condi2-2}
\end{eqnarray}%
Conditions (\ref{condi2-1}) and (\ref{condi2-2}) represent two conditions
for region $t$ and two conditions for region $d$, each involving the
corresponding set of four potentials. For example, in region $t$ condition (%
\ref{condi2-1}) is a simple algebraic equation because the potentials do not
depend on $P$ explicitly (a possible physical choice on the CMF), while
condition (\ref{condi2-2}) is a partial differential equation. We have
completed the determining equations for the potentials, which, for example,
in region $t$ is composed of (\ref{firstwo-1}) and (\ref{firstwo-2})
together with the $t$ sector of Eqs. (\ref{condi2-1}) and (\ref{condi2-2}).
At this point we notice another reason to include only four ghost
potentials, as we found four determining equations to be satisfied [Eqs. (%
\ref{firstwo-1}), (\ref{firstwo-2}), (\ref{condi2-1}) and (\ref{condi2-2})
]. The solution to these determining partial differential equations should
determine the potentials in the CMF. This solution is elaborate and still
involves arbitrary initial functions of $r$, which must be determined
numerically, which we discuss elsewhere \cite{PRLwhere}. \ In this work we
calculate the Hamiltonian orbits directly with an independent numerical
method.

\ Finally we notice a symmetry relating region $d$ of case $a$ to region $t$
of case $b$ of the equal-mass case: As the direction of time in the CMF is
arbitrary and the particles are identical, the Lagrangian for the $d$ branch
of case $a$ must be equal to the Lagrangian for the $t$ branch of case $b$
with particles exchanged and vice-versa, which implies
\begin{eqnarray}
\alpha _{t} &=&-\beta _{d},\qquad \beta _{t}=-\alpha _{d},
\label{time-and-exchange} \\
\phi _{t} &=&-\phi _{d},\qquad V_{t}=-V_{d},  \nonumber
\end{eqnarray}%
A consequence of Eq. (\ref{time-and-exchange}) is that $E_{a}^{t}=E_{b}^{d}$
and $E_{a}^{d}=E_{b}^{t}$ as well as $P_{a}^{t}=P_{b}^{d}$ and $%
P_{a}^{d}=P_{b}^{t}$. Because $\phi $ and $V$ are arbitrarily defined up to
gauge constants, and as $\phi $ and $V$ enter with a plus sign in case $a$
and with a minus sign in case $b$, we can also choose $E_{a}^{t}=E_{b}^{t}=E$
and $P_{a}^{t}=P_{b}^{t}=P$, such that one can use a common value for all
the energies and a common value for all momenta, throughout the four
combinations of region and case. We henceforth indicate energies simply by $%
E $ and momenta by $P$.\bigskip

\bigskip

\section{\protect\bigskip Equation of matching for the equal-mass case}

\bigskip In this section we introduce a simpler description in terms of two
simple functions $s(r,E)$ and $F(r,E)$ that are immediately accessible
numerically. For example the $d$ sector of (\ref{condi2-1}) and (\ref%
{condi2-2}) is studied by defining $\partial p_{a}^{d}/\partial P$ , $%
\partial p_{a}^{d}/\partial E$, $\partial p_{b}^{d}/\partial P$, and $%
\partial p_{b}^{d}/\partial E$ in terms of $s(r,E)$ and $F(r,E)$ as
\begin{eqnarray}
\frac{\partial p_{a}^{d}}{\partial P} &=&\frac{\partial p_{b}^{d}}{\partial P%
}\equiv -\frac{\cosh [s(r,E)]}{\sinh [s(r,E)]},  \label{defiFs-1} \\
\frac{\partial p_{a}^{d}}{\partial E} &=&\frac{\partial p_{b}^{d}}{\partial E%
}\equiv \frac{F(r,E)}{\sinh [s(r,E)]}.  \label{defiFs-2}
\end{eqnarray}%
For branch $t,$ the consequences of definitions (\ref{defiFs-1}) and (\ref%
{defiFs-2}) and the symmetry relations of (\ref{time-and-exchange}) are: (i)
that branch $t$ involves the same function $F(r,E)$ of branch $d$ and (ii)
that branch $t$ involves the function $s(r,E)$ of branch $d$ with a change
of sign. The intuitive picture is that $\pm s(r,E)$ and $F(r,E)$ describe
both case $a$ and case $b,$ exchanging branches in the same case replaces $%
s(r,E)$ by $-s(r,E)$, while exchanging case for the same branch leaves
functions $s(r,E)$ and $F(r,E)$ unchanged. In the following we drop the
dependence on $E$ of the functions for brevity.

Now we must impose that the same orbit is a solution of both Eqs. (\ref%
{hamia}) and (\ref{hamib}), which demands that the $d$ portion of the $a$
orbit of particle $1$ should coincide with a \emph{piece} of the $t$ branch
of particle $1$ in case $b$ (see Fig. 1). Notice that this is not the
one-to-one branch correspondence of the symmetry considerations of Sec. III,
and we stress the word \emph{piece}, because the branches are changed at
different points, as can be seen from Fig. 1. We shall henceforth drop the
subscript notation, and simply write $s_{a}$ and $s_{b},$ meaning the plus
or the minus branch of the function $s(r)$, wherever it applies. We can use
Eqs. (\ref{tablea}) , (\ref{tableb}) and (\ref{defiFs-1}), (\ref{defiFs-2})
to express the differentials of the particle-$1$ coordinates with type $a$
foliation in terms of $r_{a}$ and $s_{a}=\pm s(r_{a}):$%
\begin{eqnarray}
dt_{1a}+dx_{1a} &=&\frac{-1}{\sinh (s_{a})}F(r_{a})dr_{a},  \label{1a} \\
dt_{1a}-dx_{1a} &=&-\frac{\exp (s_{a})}{\sinh (s_{a})}dr_{a},  \nonumber
\end{eqnarray}%
and with type $b,$%
\begin{eqnarray}
dt_{1b}+dx_{1b} &=&\frac{\exp (s_{b})}{\sinh (s_{b})}dr_{b},  \label{1b} \\
dt_{1b}-dx_{1b} &=&\frac{1}{\sinh (s_{b})}F(r_{b})dr_{b}.  \nonumber
\end{eqnarray}%
At this point it is convenient to introduce still another function: the
velocity function of particle $1$, which must be the same in the
corresponding branches of each case. We define it in case $a$ by $%
v_{1a}(s_{a},r_{a})=\tanh (\Phi )$ and in case $b$ by $v_{1b}(s_{b},r_{b})%
\equiv \tanh (\Phi )$, which yields%
\begin{equation}
\exp (2\Phi )\equiv \exp (-s_{a})F(r_{a})=\exp (s_{b})F^{-1}(r_{b}).
\label{defvel}
\end{equation}%
The first condition of matching for the trajectory of particle $1$ as
described by the two foliations, requires that the velocities be the same, $%
v_{1a}(s_{a},r_{a})=v_{1b}(s_{b},r_{b})$, resulting in%
\begin{equation}
\exp (s_{a})\exp (s_{b})=F(r_{a})F(r_{b}),  \label{paralelism}
\end{equation}%
a rearrangement of Eq. (\ref{defvel}). It is important to stress that
differently from the symmetry conditions, in this condition $r_{a}$ and $%
r_{b}$ are not equal, but rather for every pair $(s_{a},r_{a})$ we should be
able to find a pair $(s_{b},r_{b})$ such that (\ref{paralelism}) is
satisfied. Fig. 2 illustrates yet another symmetry special for the
equal-mass case: while particle $1$ has a velocity angle $\Phi _{t}(r_{a})$
(event $v_{1a}^{t}$\ in Fig. 2), particle $2$ has a velocity angle of $\Phi
_{d}(r_{a})$ , the same velocity particle $1$ had in the past at the first
time that $r_{a}=r$ \ [this symmetry reads $v_{2a}^{t}(r)=v_{1a}^{d}(r)$].
Notice that there are two points along the orbit where the advanced distance
assumes a given value, one in the $t$ branch where the velocity angle of
particle $1$ is $\Phi _{t}(r_{a})$ and one in the $d$ branch with velocity
angle $\Phi _{d}(r_{a})$, as illustrated in Fig. 2. With the understanding
that these two branches must be produced with opposite signs for the
function $s(r)$, equation (\ref{defvel}) implies that%
\begin{equation}
\exp (\Phi _{d})\exp (\Phi _{t})=F(r_{a}),  \label{delayconstraint}
\end{equation}%
which in turn shows that $F(r)$ is determined by past data only, namely the
function $\Phi _{d}(r_{a}).$ Another consequence of Eq. (\ref%
{delayconstraint}) is that $F(\infty )=1,$ as $\Phi _{d}(\infty )=-\Phi
_{t}(\infty )$, the asymptotic boundary condition on the CMF. Once the orbit
is described by two differentials, there is another condition for the orbits
to be parallel at all times, which is most easily expressed by equating the
relativistic proper time of particle $1$ in the two foliations:%
\begin{equation}
(d\tau _{1})^{2}=\frac{\exp (s_{a})}{\sinh ^{2}(s_{a})}F(r_{a})(dr_{a})^{2}=%
\frac{\exp (s_{b})}{\sinh ^{2}(s_{b})}F(r_{b})(dr_{b})^{2}.  \label{taueq}
\end{equation}%
From the above we can derive differential equations for the motion of $r_{a}$
and $r_{b}:$%
\begin{eqnarray}
\frac{dr_{a}}{d\tau _{1}} &=&-\frac{\exp (-s_{a}/2)\sinh (s_{a})}{\sqrt{%
F(r_{a})}},  \label{motionr} \\
\frac{dr_{b}}{d\tau _{1}} &=&\frac{\exp (-s_{b}/2)\sinh (s_{b})}{\sqrt{%
F(r_{b})}}.  \nonumber
\end{eqnarray}%
Notice that we have used opposite signs for the evolution of $r_{a}$ and $%
r_{b\text{ }}$, the only sensible choice. Equation (\ref{motionr}) describes
a decrease of $r_{a}$ and $\ r_{b}$ at large distances if $s_{a}>0$ and $%
s_{b}<0$ (ingoing asymptotics) and an increase of $r_{a}$ and $r_{b}$ at
large distances when $s_{a}<0$ and $s_{b}>0$ (outgoing asymptotics). While
asymptotically $s_{a}$ and $s_{b}$ must have opposite signs, they do not
change sign at the same point and, in particular, in the turning region of
particle $1$ they have the same sign, as illustrated in Fig. 1. We can also
eliminate $s_{a\text{ }}$and $s_{b}$ in favor of $\Phi $ from Eq. (\ref%
{motionr}), resulting in
\begin{eqnarray}
\frac{dr_{a}}{d\tau _{1}} &=&-\frac{1}{2}[\exp (-\Phi )-\frac{\exp (3\Phi )}{%
F^{2}(r_{a})}],  \label{finalrarb} \\
\frac{dr_{b}}{d\tau _{1}} &=&\frac{1}{2}[\exp (\Phi )-\frac{\exp (-3\Phi )}{%
F^{2}(r_{b})}].  \nonumber
\end{eqnarray}%
To close the dynamical system of matching we need an equation for the
variable $\Phi $; which is provided by the Wheeler-Feynman equation of
motion (\ref{eqmotion}). To obtain a local equation, we write Eq. (\ref%
{eqmotion}) using a combination of type $a$ and type $b$ foliations in the
following way: whenever we need the advanced position of particle 2, we
write it using type $a$ foliation (as particle $2$ is naturally in the
future light-cone), while the retarded position of particle $2$ is simply
written with type $b$ foliation (where particle $2$ is naturally in the past
light-cone). The usefulness of the variable $\Phi $ is discovered when Eq. (%
\ref{eqmotion}) is written in terms of $\Phi $ and the proper time of
particle $1$, which yields simply (recall that we are using $e=1$)%
\begin{equation}
\frac{d\Phi }{d\tau _{1}}=\frac{1}{2}\{\frac{\exp (2\Phi )}{%
r_{a}^{2}F^{2}(r_{a})}+\frac{\exp (-2\Phi )}{r_{b}^{2}F^{2}(r_{b})}\}.
\label{finalphi}
\end{equation}%
Equations (\ref{finalrarb}) and (\ref{finalphi}) constitute the complete
ordinary differential equation (ODE) to describe the matching for the orbit
of particle $1$. By now we have turned equation (\ref{eqmotion}) upside down
and used all the symmetries, and the resulting Eqs. (\ref{finalrarb}) and (%
\ref{finalphi}) are much simpler to solve than equation Eq. (\ref{eqmotion}%
). \ Rigorously speaking we now have a delay-only equation, as $F(r)$
depends only on past data via (\ref{delayconstraint}). This should be
contrasted with Eq. (\ref{eqmotion}), a neutral-delay-advance equation with
infinite lags. To solve Eqs. (\ref{finalrarb}) and (\ref{finalphi}) one
needs to postulate an arbitrary positive function $F(r)$ with a given
asymptotic form $F(\infty )=1$, and solve the resulting ODE. For
self-consistency the ghost function $F(r)$ must be chosen such that the
orbit of particle $2$ is the same even function as that of particle $1$, by
definition of the CMF. This functional problem is solved numerically in the
following section.

Last, as an illustration, we outline how one can express the potentials in
the CMF in terms of the numerically accessible functions $s(r,E)$ and $%
F(s,E) $ along the simplest analytical solution of Eqs. (\ref{firstwo-1}), (%
\ref{firstwo-2}), (\ref{condi2-1}) and (\ref{condi2-2}). As the potentials
in the CMF do not depend on $P$, Eq. (\ref{defiFs-1}) is algebraic and can
be solved simply in both case $a$ and case $b.$ One possible solution is%
\begin{equation}
P_{a}+V+\frac{e^{2}}{r}=\frac{[2(1+\alpha )(1+\beta )\cosh (s)-(1+\alpha
)^{2}-(1+\beta )^{2}]}{4(E+\phi )},  \label{eq47}
\end{equation}

\bigskip and%
\begin{equation}
P_{b}-V+\frac{e^{2}}{r}=\frac{[2(1-\alpha )(1-\beta )\cosh (s)-(1-\alpha
)^{2}-(1-\beta )^{2}]}{4(E-\phi )}.  \label{eq48}
\end{equation}%
For case $a$ the equations of motion for $\xi _{1}$ and $\xi _{2}$ derived
from Eqs. ( \ref{hamia}) and (\ref{firstwo-1}) are
\begin{equation}
\frac{d\xi _{1}}{d\zeta }=\frac{\exp (s)}{F(r)}=\frac{(1+\alpha )^{2}}{%
4(p_{1}+\frac{1}{2}V+\frac{e^{2}}{2r})^{2}},  \label{Eq49}
\end{equation}%
and
\begin{equation}
\frac{d\xi _{2}}{d\zeta }=\frac{\exp (-s)}{F(r)}=\frac{(1+\beta )^{2}}{%
4(p_{2}+\frac{1}{2}V+\frac{e^{2}}{2r})^{2}}.  \label{Eq50}
\end{equation}%
We now take the most physically sensible square root of Eqs. (\ref{Eq49})
and (\ref{Eq50}) and substitute into Eq. (\ref{hamia}), yielding

\[
E_{a}^{d}=-\phi _{d}+\frac{1}{2\sqrt{F}}[(1+\alpha _{d})\exp (s/2)-(1+\beta
_{d})\exp (-s/2)].
\]%
In the same way, for case $b$, we obtain after a choice of sign for the
square root%
\[
E_{b}^{d}=\phi _{d}-\frac{1}{2\sqrt{F}}[(1-\alpha _{d})\exp (s/2)-(1-\beta
_{d})\exp (-s/2)].
\]%
As discussed below Eq.(\ref{time-and-exchange}), we henceforth set $%
E_{b}^{d}=E_{a}^{d}$ $=E$ and $P_{b}^{d}=P_{a}^{d}=P$, which defines the
explicit solution for $\phi _{d}\,$, $V_{d}$, $\alpha _{d}$ and $\beta _{d}$
as%
\begin{eqnarray}
\phi _{d} &=&\frac{\sinh (s/2)}{\sqrt{F}},  \label{phiV} \\
V_{d} &=&\sqrt{F}\sinh (s/2).  \nonumber
\end{eqnarray}%
The resulting equations for $\alpha _{d}$ and $\beta _{d}$ are

\begin{eqnarray}
\alpha _{d}\exp (s/2)-\beta _{d}\exp (-s/2) &=&2E\sqrt{F},  \label{alphabeta}
\\
\alpha _{d}\exp (-s/2)-\beta _{d}\exp (s/2) &=&-\frac{2(P+\frac{e^{2}}{r})}{%
\sqrt{F}},  \nonumber
\end{eqnarray}%
where $E\equiv \frac{1}{2}(E_{a}^{d}+E_{b}^{d})$ and $P=\frac{1}{2}%
(P_{a}^{d}+P_{b}^{d})$ are the sole two physically meaningful constants of
the problem. At the shortest light-cone distance $r_{o}$ , which happens at $%
s(r_{o})=0$ , the determinant of the linear system (\ref{alphabeta})
vanishes and poses the following condition involving $E$ and $P$%
\begin{equation}
E=-\frac{(P+\frac{e^{2}}{r_{o}})}{F(r_{o})}.  \label{relEP}
\end{equation}%
For $r>r_{o}$ one has $s(r)\neq 0$ and (\ref{alphabeta}) can be solved for $%
\alpha _{d}$ and $\beta _{d}$ yielding%
\begin{eqnarray}
\alpha _{d} &=&\frac{[EF(r)\exp (s/2)+(P+\frac{e^{2}}{r})\exp (-s/2)]}{\sqrt{%
F(r)}\sinh (s)},  \label{solalphbet} \\
\beta _{d} &=&\frac{[EF(r)\exp (-s/2)+(P+\frac{e^{2}}{r})\exp (s/2)]}{\sqrt{%
F(r)}\sinh (s)}.  \nonumber
\end{eqnarray}%
In the CMF $P$ is a function of $E$, so that the potentials given by Eqs. (%
\ref{phiV}) and (\ref{solalphbet}) depend only on $E$ and $r$. We warn the
reader that this simplest type of solution of Eqs. (\ref{firstwo-1}), (\ref%
{firstwo-2}), (\ref{condi2-1}) and (\ref{condi2-2}) does not correspond to
the low-energy orbits of the repulsive case studied here numerically in the
following section, and was included as an illustration of how $s(r,E)$ and $%
F(r,E)$ can determine the potentials in the simplest possible way. There are
several other solutions to Eqs. (\ref{firstwo-1}), (\ref{firstwo-2}), (\ref%
{condi2-1}) and (\ref{condi2-2}), and the detailed consideration of all
cases and energy ranges requires further elaborate analytical work and
extensive comparison with numerics, and will be published elsewhere.

\section{\protect\bigskip Numerical Integration: the steepest-descent method}

\bigskip In Sec. IV\ we saw that describing the same particle $1$ in both
foliations results in Eqs. (\ref{finalrarb}) and (\ref{finalphi}), involving
the single unknown ghost function $F(r)$. The above discussion suggests the
following simple self-consistent method to obtain the symmetric solution of
1D-WF2B in the equal-mass case in the CMF:\ We start by postulating the
functional form of $F(r)$, which must go to $1$ at large distances, as noted
below Eq. (\ref{delayconstraint}). For the following numerical work we use
up to $18$ arbitrary coefficients to approximate $F(r)$ by a truncated power
series,
\begin{equation}
F(r)=1-\sum_{n=1}^{n=18}\frac{k_{n}}{r^{n}},  \label{expreF}
\end{equation}%
which has the desired asymptotic form. After we assume given values for the $%
k_{n}$, the main equations (\ref{finalrarb}) and (\ref{finalphi}) yield a
simple initial value ODE problem. The integration can be carried\ out from
the turning point, where $\tau _{1}=0,$ $\Phi =0$ and $%
r_{a}=r_{b}=r_{c}>r_{o}$. \ (the given functional form of $F(r)$ and $r_{c}$
determine all the subsequent dynamics). Notice that this integration
automatically produces a time-reversible orbit for particle 1: It can be
seen by inspection of \ Eqs. (\ref{finalrarb}) and (\ref{finalphi}) that
replacing $\tau _{1}$ by $-\tau _{1}$ \ and $\Phi $ by $-\Phi $ simply
exchanges $r_{a}$ and $r_{b},$ such that $r_{a}(-\tau _{1})$ $=$ $r_{b}(\tau
_{1})$, a consequence of the symmetry imposed. When we start particle $1$ at
the turning point, $\Phi =0$, particle $2$ is described in type $a$
parametrization at the advanced point $(x_{2a},t_{2a})=(-r_{c},r_{c})$,
while with type $b$ parametrization particle $2$ is at the retarded position
$(x_{2a},t_{2a})=(-r_{c},-r_{c}).$ As illustrated in Fig. 3, for a generic
choice of $F(r)$ the future of type $b$ trajectory of particle $2$ will not
match the type $a$ trajectory of the same particle $2$, which starts ahead
of case $b$, and the scheme produces two different orbits for particle $2$,
which is an absurd. It is necessary to adjust the function $F(r)$ precisely
to obtain a single trajectory for particle $2$, and it is nice to observe
that the asymptotic condition $F(\infty )=1$ guarantees the asymptotic
velocity of particle $2$ to be the same in both foliations, so if we adjust
the orbits to overlap in the turning region, they become close everywhere.

Our numerical method produces two trajectories for particle $2$ from each
set of $k_{n}$, by direct numerical integration of the main equations (\ref%
{finalrarb}) and (\ref{finalphi}) accompanied by the driven equations for
the trajectories:$\,(\frac{dx_{2a\,}}{dr_{a}})$ $\,,(\,\frac{dt_{2a}}{dr_{a}}%
)\,\,,\,(\frac{dx_{2b}}{dr_{b}})\,$\ and $\,(\frac{dt_{2b}}{dr_{b}})$ as
determined by Eqs. (\ref{tablea}) and (\ref{tableb}). We calculate the
trajectories numerically by using a 9/8 embedded Runge-Kutta pair. In
general, two different trajectories are obtained for particle $2$, as
illustrated in Fig. 3 and we calculate numerically the average squared
deviation of the two trajectories over a grid of positions%
\begin{equation}
A(k)=\sqrt{\frac{1}{N}\sum_{i=1}^{N}(t_{2a}(x_{2i})-t_{2b}(x_{2i}))^{2}.}
\label{squaredev}
\end{equation}%
After that we implement a steepest-descent search in the $18$-dimensional
parameter space governed by the quenching equation $dk_{n}/ds=-\partial
A/\partial k_{n}$ until it finds a minimum value for the squared deviation
of \ Eq. (\ref{squaredev}) (see Ref. \cite{Vishal} for an analogous
numerical quenching procedure).

In Fig. 4 we put the converged trajectory of particle $1$ superposed to the
reflected trajectory of particle $2\,,$ for velocities $v/c=0.46$\thinspace
, $\,v/c=0.54$ and $v/c=0.71$. Notice that the trajectories coincide
perfectly, indicating that the quenching search satisfied all the symmetry
relations and thus found a consistent solution. As the solution is
self-consistent, we can not set the asymptotic velocity directly, and we
determine a final low velocity by using an initial condition with $r_{c}\gg
1 $, while a large asymptotic velocity is achieved by using $r_{c}\approx 1$
(one classical electronic radius). We start the numerical work at low
velocities, by setting a large value of $r_{c}$, which results in a small
asymptotic velocity. After that, we decrease the value of $r_{c}$ and give
the formerly determined solution as seed to the quenching method, which
converges much faster as there must be a solution in the neighborhood of the
seed.

In Table 1 we list the final velocities $v/c$ as a function of the initial
condition $r_{c}$. From the numerically converged trajectories we calculate
the minimum distance $r_{o}$ and Table 1 shows $r_{o}/r_{c}.$ We observe
that some coefficients $k_{n}$ converge to a value below the numerical
precision of $10^{-14}$ such that only the first $N$ coefficients are
significant to the numerical precision. This number increases with the
asymptotic velocity, as can be seen from Table 1.

In Fig. 5 we plot the trajectory of $v/c=0.80$ , which took twelve hours of
numerical quenching to converge and still one can observe a slight mismatch
of the orbits in the turning region, indicating the slowness of the
convergence process. For this case, the numerical $19th$ coefficient is
still important, indicating that our basis is failing to converge to the
solution, which is suggestive that something physical is happening above $%
v/c=0.71$, maybe we are even loosing the twice-monotonic property (according
to Appendix 1 this could happen at any point above $v/c=0.33$). There might
also be another trajectory in the neighborhood, which interferes with the
convergence, and last, at high energies the form of $F(r)$ as given by Eq. ( %
\ref{expreF}) \ becomes too singular at the collision and we need a
regularized numerical method. Further numerical studies with a regularized
numerical method are needed to determine if some special bifurcation is
happening to the orbit above $v/c=0.71$. In this high-velocity region the
functional problem posed on $F(r)$ might not have a unique solution, and for
relativistic velocities it is likely that it does not. The symmetric
solution was actually proved to be unique only up to a small velocity \cite%
{Driver1}, and we managed to go much above the low limit set by Driver in
Ref. \cite{Driver1}. In Fig. 6 we plot $F(r)$ versus $(r_{o}/r)$ for the
asymptotic velocities $v/c=0.46$, $v/c=0.54$ and $v/c=0.80$. Notice that the
functional form of $F(r)$ is approximately a linear function of $(r_{o}/r)$
at low velocities, but at larger velocities it becomes highly convoluted.

As a test for the existence of other types of orbits, we integrated Eqs. (%
\ref{finalrarb}) and (\ref{finalphi}) using two completely general (and
possibly different) functions $F_{a}(r_{a})$ and $F_{b}(r_{b}),$ each
defined by an independent truncated power series like%
\begin{equation}
F_{a,b}(r)=k_{o}^{a,b}-\sum_{n=1}^{n=9}\frac{k_{n}^{a,b}}{r^{n}}.
\label{expreFab}
\end{equation}%
Notice that the saturation value is not anymore $k_{o}^{a,b}=1$ like in Eq. (%
\ref{expreF}), but rather a generic quenchable value in each case. The
integration procedure was started with $r_{a}\neq r_{b}$ at $\Phi =0$ for
particle $1.$ We found that the quenching method converged to functions $%
F_{a}(r_{a})$ and $F_{b}(r_{b})$ that were generally different but were
always related by the scaling discussed in Appendix 2 [see equation (\ref%
{formofF}) and text below it]. According to the discussion of Appendix 2
this is the case when the orbit is symmetric in another Lorentz frame. This
above-defined search would be capable of finding any existing orbit with the
twice-monotonic property and the fact that it always converged to
Lorentz-transformed symmetric orbits is indicative that there are no other
types of low-energy solutions.

At this point it is interesting to appreciate the big detour taken by our
numerical method to solve Eq. (\ref{eqmotion}), which should be compared to
the most straightforward way to solve a neutral-delay differential equation
like Eq. (\ref{eqmotion}), namely: postulating an initial function and
continuing the solution by use of the differential equation. The
straightforward method necessarily leads to runaways because one is never
capable of guessing the unique nonrunaway initial function, and even if one
does guess the nonrunaway condition, numerical roundoff plagues the
integration and one still gets runaways after some short time. Our numerical
method is superior in this respect precisely because it is already placed in
the nonrunaway manifold and the quenching implemented to solve the
functional problem for $F(r)$ is numerically stable, as it does not involve
extrapolation.

\

\section{\protect\bigskip Conclusions and Discussion}

In this paper we discussed the solution and Hamiltonian description of the
time-symmetric two-body problem of the action-at-a-distance electrodynamics
with repulsive interaction. Our method is closed and does not involve
expansions, only the hypothesis that the orbit is twice-monotonic was used.
We conjecture that our solution is already the general solution at low
energies, which can be argued in the following way: For a generic solution,
possibly unsymmetrical, one can always find a Lorentz frame where the
asymptotic outbound velocities of the two particles are opposite (the
outbound CMF). Now \emph{if} asymptotic data determine a unique trajectory
for low energies, that trajectory is our symmetric solution in this CMF. Of
course the symmetric solution is always a possible solution, but the
symmetry of the physics in this outbound CMF suggests that it is the only
solution for low energies, as the solution has to correspond to the
Coulombian solution, which has this property (this would actually be a
non-trivial generalization of the work of Driver and Hoag in references \cite%
{Driver1,Driver2}). With the above in mind, the Hamiltonian we derive here
is already the general order-reduced Hamiltonian for low energies. Some few
numerical experiments have suggested that the conjecture is correct at low
energies. Even the high energy solutions found in Ref. \cite{Igor-pair}
exhibit the property that the future of one particle is the past of the
other, and it looks like the numerical method in Ref. \cite{Igor-pair}
simply picked a Lorentz frame slightly off the CMF, but that requires
further investigation with a regularized numerical method.

The idea to remove the field degrees of freedom goes back to Dirac \cite%
{Dirac} and later Wheeler and Feynman planned to quantize WF2B as a means to
avoid the divergencies of QED, as in the action-at-a-distance theory the
infinite number of field degrees of freedom is absent. History says that the
famous seminar that never came from Wheeler (see Ref. \cite{Mehra}, page 97)
was due to difficulties in converting the Fokker Lagrangian (\ref{Fokker})
to the Hamiltonian form. This task is still not fully done and in this work
we took a step in that direction for the one-dimensional case at low
energies. Notice that the implicit dependence of the Hamiltonian operator is
actually convenient for an eigenvalue equation, and one could discuss a
canonical quantization procedure based on either Eq. (\ref{hamia}) or Eq. (%
\ref{hamib}), using the numerically determined potentials. Of course Wheeler
and Feynman were mainly interested in the attractive case, of greater
relevance for atomic physics and specially for the Lamb shift calculation.
The attractive problem is being published elsewhere \cite{PRLwhere}. In this
same chapter 5, page 97 of reference \ \cite{Mehra}, Feynman says that \ ` I
didn 't solve it either---a quantum theory of half-advanced half-retarded
potentials---and I worked on it for years... '. This is still an outstanding
problem today and the difficulties in casting relativistic \emph{Lagrangian}
interactions into the \emph{Hamiltonian} form are well explained in
references \cite{Tretyak,ListNaza}. The only studies we know of dealing with
the time-symmetric problem involve power expansions. We are aware of another
attempt at a Hamiltonian description of 1D-WF2B that ends up with an
infinite-dimensional Hamiltonian \cite{Igors}, such that further order
reduction is needed to select nonrunaway orbits.

Our description might seem to violate the no-interaction theorem \cite%
{Currie,Constraint}, but there are two places where we avoid it: (i) the
no-interaction theorem is an obstacle to covariant Hamiltonian description
of two interacting particles only in fully 3-dimensional motion. We are
restricted to 1-dimensional motion; (ii) The evolution parameter in our
Hamiltonians is not time but rather $\zeta $ for case $a$ or $\xi $ for case
$b$ and, therefore, the no-interaction theorem does not apply. In principle,
because time is not the evolution parameter, even in three dimensions the
no-interaction theorem would not be an obstacle to an analogous procedure,
and that is an open problem.

As regards the applied mathematics literature of delay, the theory of
delayed functional equations \cite{Diekmann,Elsgolts} is a difficult and
poorly investigated subject but it turns out that there are already a few
results worth noticing: In a paper of 1974 by Kaplan and Yorke \cite{Kaplan}
it was noticed that for some special types of delay\emph{\ }equations,
solutions can be found by searching the \emph{periodic orbits} of an
associated \emph{ordinary} differential equation. This was further
generalized in 1999 \cite{Jibin} and it was shown that for a large class of
delay equations the associated ODE turned out to be a \emph{Hamiltonian}
ODE, quite a curious result\cite{Jibin} brought up by applied mathematicians
with no relation to either quantum mechanics or Wheeler-Feynman
electrodynamics. Another set of studies of applied mathematics focuses on
the similarities of delay equations to either ODE's or extended systems \cite%
{Politi}: if the delay is small and bounded, the behavior should be
reminiscent of that of ODE's, as determined by the dimension of the
attractor in several systems with small delays, a generic class that
contains the bound states of the attractive 1D-WF2B, apart from the fact
that our system is conservative. In the limit where the delays are very
large, delay equations are found to behave like extended systems, with large
dimensional attractors, which is the generic class of the repulsive case of
1D-WF2B, where the lags are unbounded and also of the unbound states of the
attractive case.

As we mentioned in the Introduction, the action-at-a-distance
electrodynamics is capable of describing the whole of classical
electrodynamics as a limiting case, and even better, a limiting case without
the complications of mass renormalization, as demonstrated by Wheeler and
Feynman \cite{Fey-Whe}. This was actually what led Wheeler and Feynman to
the action-at-a-distance electrodynamics in the first place, but in doing
that they formulated a very complex \emph{conservative} physical theory (the
conserved energy associated with the Fokker Lagrangian is discussed, for
example, in Ref. \cite{Anderson}). It is important to stress that the
converse of the above statement is not true at all: the complex \emph{%
conservative} dynamics of the action-at-a-distance theory is not reducible
to a limiting case of Maxwell's electrodynamics (which is always a
dissipative theory because of the radiation). Relativistic
action-at-a-distance shares the conservative character with Newtonian
gravitation, and in the presence of a universe of particles, an atom
described by the action-at-a-distance theory has the possibility to behave
in a way analogous to the solar system in the Newtonian sky: \ distant solar
masses being just small perturbations, as opposed to the description a la
Maxwell, where it \emph{has} to radiate. It appears to us that the analysis
of the complex conservative dynamics of WF2B is bound to reveal interesting
new physical insights. As we have seen with this special case study, the
physical nonrunaway condition performs the magical reduction from the
infinite-dimensional dynamical system posed by the delay equation to a
finite-dimensional one, and the large body of existing understanding on
qualitative behavior of finite-dimensional vector fields should be applicable%
\cite{Diekmann}. Some results already published for systems of atomic
physics within the Darwin approximation, a low-velocity Hamiltonian
approximation to action-at-a-distance, have already revealed interesting news%
\cite{PRL,discrete}. Existing numerical methods for the relativistic case
are still short reaching \cite{Bonn} and can not reproduce the massive
numerical search performed with the Darwin approximation in \cite{discrete}.

\bigskip

\section{ Appendix 1: Proof of the twice-monotonic property}

\qquad

In this section we show that in the Coulombian limit of a low-energy orbit,
the solution of Eqs. (\ref{eqmotion}) and\ (\ref{lightcone})\ has only two
branches, one defined by $\dot{r}>0$ and another defined by $\dot{r}<0$.
Because of time-reversal symmetry, the theorem is the same for either $r_{a}$
[indicated by $q$ in Eqs. (\ref{eqmotion}) and (\ref{lightcone})] or by $%
r_{b}$ [indicated by $r$ in Eqs. (\ref{eqmotion}) and (\ref{lightcone})]. It
suffices to prove that there is only one point where $\dot{q}$ vanishes,
with $q$ defined in Eq. (\ref{lightcone})). A special version of this proof
was given in Ref. \cite{Driver1}\ along symmetric orbits of the equal-mass
case. The proof is trivial and can be done for a generic orbit of the
arbitrary-mass repulsive two-body system in the CMF: We start from the
definition of the light-cone condition for a generic CMF orbit,%
\begin{equation}
q=x_{1}(t)-x_{2}(t+q),  \label{generalcone}
\end{equation}%
where $x_{1}(t)$ represents the position of particle $1$, assumed on the
right, and $x_{2}(t)$ represents the position of particle $2$, assumed on
the left, and we have set $c=1$ . Notice that Eq. (\ref{lightcone}) is a
special case of Eq. (\ref{generalcone}) for symmetric orbits of the equal
mass case [$x_{2}(t)=-x_{1}(t)$ ]. The key observation is that because the
interaction is always repulsive, the velocity $v_{1}(t)$ of particle $1$ is
a monotonically \emph{increasing }function of time (particle $1$ is repulsed
to the right), while the velocity $v_{2}(t)$ of particle $2$ is a
monotonically \emph{decreasing} function of time (particle $2$\ is repelled
to the left). If we take the derivative of Eq. (\ref{generalcone}) respect
to $t$ and isolate $\dot{q}$ we obtain%
\begin{equation}
\dot{q}=\frac{v_{1}(t)-v_{2}(t+q)}{1+v_{2}(t+q)}.  \label{equationqdot}
\end{equation}%
For low-energy we have the bounds $|v_{2}(t+q)|<v_{2}(\infty )\ll 1$ and $%
|v_{1}(t)|<v_{1}(\infty )\ll 1$ , and therefore the denominator of \ Eq. (%
\ref{equationqdot}) is always positive. In the CMF the value of $\dot{q}$
changes sign from the inbound asymptotic region to the outbound asymptotic
region, with values onto the interval

\begin{equation}
-(\frac{v_{1}(\infty )+v_{2}(\infty )}{1+v_{2}(\infty )})<\dot{q}<(\frac{%
v_{1}(\infty )+v_{2}(\infty )}{1-v_{2}(\infty )}).  \label{asymbounds}
\end{equation}

To complete the proof we need only to notice that the sum of two
monotonically increasing functions [$v_{1}(t)$ and $-v_{2}(t+q)$ ] is also
monotonically increasing and, therefore, can only vanish once. It should be
noticed that $-v_{2}(t+q)$ is not necessarily a monotonically increasing
function of $t$ for high-velocity orbits, as%
\begin{equation}
\frac{dv_{2}(t+q)}{dt}=(1+\dot{q})\frac{dv_{2}(t+q)}{d(t+q)},  \label{monoeq}
\end{equation}%
which is the product of $(1+\dot{q})$ times a negative number [recall that $%
v_{2}(t)$ is a monotonically decreasing function of its argument and,
therefore, $-v_{2}(t)$ is a monotonically increasing function]. It can be
seen that Eq. (\ref{monoeq}) guarantees that $v_{2}(t+q)$ is an increasing
function of $t$ if $(1+\dot{q})$ is positive, which is the case for a
low-velocity orbit. A simple estimate for the first velocity where the
twice-monotonic property can fail in the equal-mass case is given by setting
$|\dot{q}|=1$ in Eq. (\ref{asymbounds}), which predicts $v(\infty )=1/3$ .

\section{Appendix 2: Covariant definition of the equal-mass case}

\bigskip In this appendix we exhibit a covariant derivation of the above
Hamiltonization procedure, which we develop only for the equal-mass case and
with the hypothesis that the orbit is twice-monotonic (proved in Ref. \cite%
{Driver1} for low energies and in Appendix 1 for the arbitrary-mass case in
the CMF). The definition of a covariant family starts from the observation
that the Lorentz transformation of a symmetric orbit has the property that
the future of particle $1$ is the past of particle $2$ and vice versa. We
define the relativistic symmetric family of orbits, henceforth called RSF,
as the family of orbits with the property that\emph{\ if} $%
x_{1}(t_{1})+x_{2}(t_{2})=0$, then it follows that $t_{1}+t_{2}=0$. \ It is
easy to verify that RSF is a Lorentz invariant family of orbits and also
that any Lorentz transformation of a symmetric orbit belongs to RSF (but, in
principle, these should not exhaust the RSF: there could be other types of
orbits). A generic orbit of the RSF is represented in Fig. 7, where we
illustrate the time-reversal and exchange symmetry. The above definition
implies that the future of particle $1$ is the past of particle $2$ in the
RSF. Inside the RSF, by use of the time-reversal operation $\zeta
_{1}\rightarrow -\xi _{2}$, $\zeta _{2}\rightarrow -\xi _{1}$, we can prove
the same identities (\ref{defphi}), (\ref{defV}) and an equivalent form of
Eq. (\ref{defab}), relating particle $1$ to particle $2$, (only that in case
$b$ the role of $\alpha $ and $\beta $ is exchanged), and these in turn lead
to the same type of general ghost Lagrangian to describe a twice-monotonic
orbit.

The relativistic condition that the future of one particle is the past of
the other implies that the solution of WF2B inside the RSF must have the
following branch correspondences: $v_{1a}^{t,d}(r_{a})=v_{2a}^{d,t}(r_{a})$
and $v_{1b}^{t,d}(r_{b})=v_{2b}^{d,t}(r_{b})$ (see Fig. 7). These conditions
can be seen with the help of (\ref{tablea}) and (\ref{tableb}) to be
equivalent to the four conditions
\begin{eqnarray}
\frac{\partial p_{a,b}^{t,d}}{\partial P} &=&-\frac{\partial p_{a,b}^{d,t}}{%
\partial P},  \label{condi0} \\
\frac{\partial p_{a,b}^{t,d}}{\partial E} &=&-\frac{\partial p_{a,b}^{d,t}}{%
\partial E}.  \nonumber
\end{eqnarray}%
The above conditions imply that the orbit is defined by four different
functions $\pm s_{a}(r),$ $\pm s_{b},F_{a}(r)$ and $F_{b}(r)$ [see
definition (\ref{defiFs-1}) and (\ref{defiFs-2}) ]. To verify that condition
(\ref{condi0}) is relativistically invariant, let us suppose that we tried
to describe the orbit from another Lorentz frame, with boost parameter $w$.
\ If the orbit is twice-monotonic, it can be described in case $a$ with a
Lagrangian $\bar{L}_{a}$ of the same type of (\ref{Lagrangian-a}) , and a
Hamiltonian of type (\ref{hamia}), and the new coordinates $\bar{x}_{1a},%
\bar{t}_{1a},$ $\bar{x}_{1b},\bar{t}_{1b}$ must be obtained by a simple
Lorentz transformation with boost parameter $w.$ Imposing this condition on
the explicit solution (\ref{tablea}), and noticing that the advanced
light-cone distance in the new frame relates to the old one by $d\bar{r}_{a}=%
\sqrt{(1+w)/(1-w)}dr_{a},$ we obtain
\begin{eqnarray}
\frac{\partial \bar{p}_{a}}{\partial E} &=&\frac{(1-w)}{(1+w)}\frac{\partial
p_{a}}{\partial E},  \label{Lorentzsymmetry} \\
\frac{\partial \bar{p}_{a}}{\partial P} &=&\frac{\partial p_{a}}{\partial P},
\nonumber
\end{eqnarray}%
which, besides showing that Eq. (\ref{condi0}) is frame independent , also
shows that $s_{a}(r)$ as defined by Eq. (\ref{defiFs-1}) is a Lorentz
scalar:
\begin{equation}
\bar{s}_{a}(\bar{r}_{a})=s_{a}(\lambda _{a}\bar{r}_{a}),  \label{formofs}
\end{equation}%
and that $F_{a}(r_{a})$ [as defined by Eq. (\ref{defiFs-2}) ] transforms
like
\begin{equation}
\bar{F}_{a}(\bar{r}_{a})=\lambda _{a}^{2}F(\lambda _{a}\bar{r}_{a}),
\label{formofF}
\end{equation}%
with $\lambda _{a}=\sqrt{\frac{1-w}{1+w}}.$ Case $b$ transforms in the same
way with $\lambda _{b}=1/\lambda _{a}.$ This last equation allows us to
express the Hamiltonian in any frame by use of the CMF form of $F(r)$ and a
rescaling depending on the boost parameter, which can be determined by
asymptotic data.

Last we show that the action of a Lorentz transformation on Hamiltonian (\ref%
{hamia}) is a canonical transformation: It is easy to verify with the help
of Eq. ( \ref{defining}\ ) that a Lorentz transformation simply rescales the
coordinates $\xi _{1}$ and $\xi _{2}$ to $\bar{\xi}_{1}=\frac{1}{\lambda }%
\xi _{1}$and $\bar{\xi}_{2}=\frac{1}{\lambda }\xi _{2}$ with $\lambda $ $=%
\sqrt{\frac{1-w}{1+w}}.$ To complete the change with a canonical
transformation one must scale the momenta with the inverse factor, $\bar{p}%
_{1}=\lambda p_{1}$ and $\bar{p}_{2}=\lambda p_{2}.$ By this canonical
transformation the transformed Hamiltonian is
\begin{equation}
\bar{H}_{a}=\frac{-\lambda }{4}\{\frac{\bar{M}_{1a}^{2}(\bar{r}_{a})}{(\bar{p%
}_{1}+\frac{1}{2}\bar{V}+\frac{e^{2}}{|\bar{\xi}_{1}-\bar{\xi}_{2}|})}+\frac{%
\bar{M}_{2a}^{2}(\bar{r}_{a})}{(\bar{p}_{2}+\frac{1}{2}\bar{V}+\frac{e^{2}}{|%
\bar{\xi}_{1}-\bar{\xi}_{2}|})}\}-\lambda \bar{\phi}(\bar{r}_{a}),
\label{canoHami}
\end{equation}%
where\ $\bar{\phi}(\bar{r}_{a})=\frac{1}{\lambda }\phi (r_{a})$, $\bar{V}(%
\bar{r}_{a})=\lambda V(r_{a})$ , $\bar{M}_{1a}^{2}(\bar{r}%
_{a})=M_{1a}^{2}(r_{a})$, and $\bar{M}_{2a}^{2}(\bar{r}%
_{a})=M_{2a}^{2}(r_{a})$ . Notice that Hamiltonian $\bar{H}_{a}$ picked a
multiplicative factor of $\lambda $ and if we also perform a change to the
natural evolution parameter $\bar{\zeta}=\lambda \zeta $ of the new Lorentz
frame, it compensates exactly for that factor, going back to the form (\ref%
{hamia}) , the same form for all Lorentz frames.

Acknowledgements: E. B. Hollander acknowledges a Fapesp PhD scholarship,
proc. 99/08316-8 and J. De Luca acknowledges CNPQ, Brazil. We thank R.
Napolitano for a careful reading of the manuscript.

\bigskip

\bigskip

\bigskip

\section{\protect\bigskip Figure Captions}

Fig. 1: Particle trajectories in the CMF for $m_{1}\neq m_{2}$ in the $(x,t)$
plane, arbitrary units with $c=1$ and $e^{2}=mc^{2}$. Particle $1$: case $a$
indicated in branches: $1a^{t}$(solid inner line on right) and $1a^{d}$
(hatched inner line on right); case $b$ indicated in branches $1b^{d}$%
(hatched outer line on right) and $1b^{t}$ (solid outer line on right). \
For clarity we indicate the type $a$ and type $b$ orbits of particle $1$ as
separate curves, but there is just one orbit for particle $1.$ Trajectory of
particle $2$ is represented by the solid line on left, branches are not
indicated. The geometric distance in the $(x,t)$ plane between two points in
light-cone condition is $\sqrt{2}$ $r$, with $r$ being the spacial distance,
we dropped the $\sqrt{2}$ factor and indicated simply $r$.

Fig. 2: \ Particle trajectories in the CMF for $m_{1}=m_{2}$ in the $(x,t)$
plane, arbitrary units with $c=1$ and $e^{2}=mc^{2}$. Indicated is the
minimum distance $r_{o}$ and the velocity of particle $1$ in each branch
corresponding to the same distance $r=r_{a}^{d}=r_{a}^{t}>r_{o}$. Event $%
v_{1a}^{d}$ is the time-reversed point of event $v_{1b}^{d}$ and, because of
that, $r_{a}^{d}=r_{b}^{d}=r.$ A special symmetry of the equal- mass case:
event $v_{2a}^{t}$ is obtained by particle exchange and time reversal from $%
v_{1a}^{d},$ which implies $r_{a}^{t}=r_{b}^{d}=r.$ All branches of the
trajectory of particle $1$ are indicated, and omitted for particle $2.$
Notice that we indicated the spacial distance in light-cone, which is $\sqrt{%
2}$ times the geometric distance in the $(x,t)$ plane.

Fig. 3: \ Matching the trajectory of particle $1$ for a generic $F(r):$
Particle $2$ starts from the past in case $b$ and, unless $F(r)$ is chosen
precisely, its future does not coincide with the trajectory of particle $2$
of case $a.$ Arbitrary units with $c=1$ and $e^{2}=mc^{2}$.

Fig. 4: \ Numerically determined trajectories in the CMF for $m_{1}=m_{2}$
in the $(x,t)$ plane; units with $c=1$ and $e^{2}=mc^{2}$. Three different
symmetric trajectories found by the steepest-descent method; the orbit of
particle $2$ is reflected and superposed onto that of particle $1$ to show
the agreement: $v/c=0.46,\,r_{c}=4.7$; $v/c=0.54,\quad r_{c}=3.7$ ; and $%
v/c=0.71$ and $r_{c}=2.97$ .

Fig. 5: Numerically determined trajectories in the CMF for $m_{1}=m_{2}$ in
the $(x,t)$ plane; units with $c=1$ and $e^{2}=mc^{2}$. The orbit of
particle $2$ is reflected and superposed to that of particle $1$, for a
highly relativistic case of $v/c=0.8$ . Notice the slight mismatch of the
two orbits, due to failure of convergence of the series.

Fig. 6: \ Numerically determined functions $F(r)$, rescaled as a function of
$(r_{o}/r)$; both quantities are dimensionless. Notice that at low
velocities $F(r)$ is well approximated by the first two terms of the
series,\ $F(r)\sim 1-\frac{k_{1}}{r}$ , but at larger velocities $F(r)$
becomes highly convoluted.

Fig. 7: A typical trajectory of the RSF in a generic Lorentz frame for the
equal mass case in the $(x,t)$ plane; units with $c=1$ and $e^{2}=mc^{2}$.
The only symmetry is that the future of particle $1$ is the past of particle
$2.$

Table 1: Numerically calculated asymptotic velocities $(v/c)$, minimal radii
$(r_{o}/r_{c})$ and number $(N)$ of significant terms of the $F(r)$ series
as a function of \ the critical initial distance $r_{c}$ at $\Phi =0.$ Units
of energy are set by $c=1$ and $e^{2}=mc^{2}$.

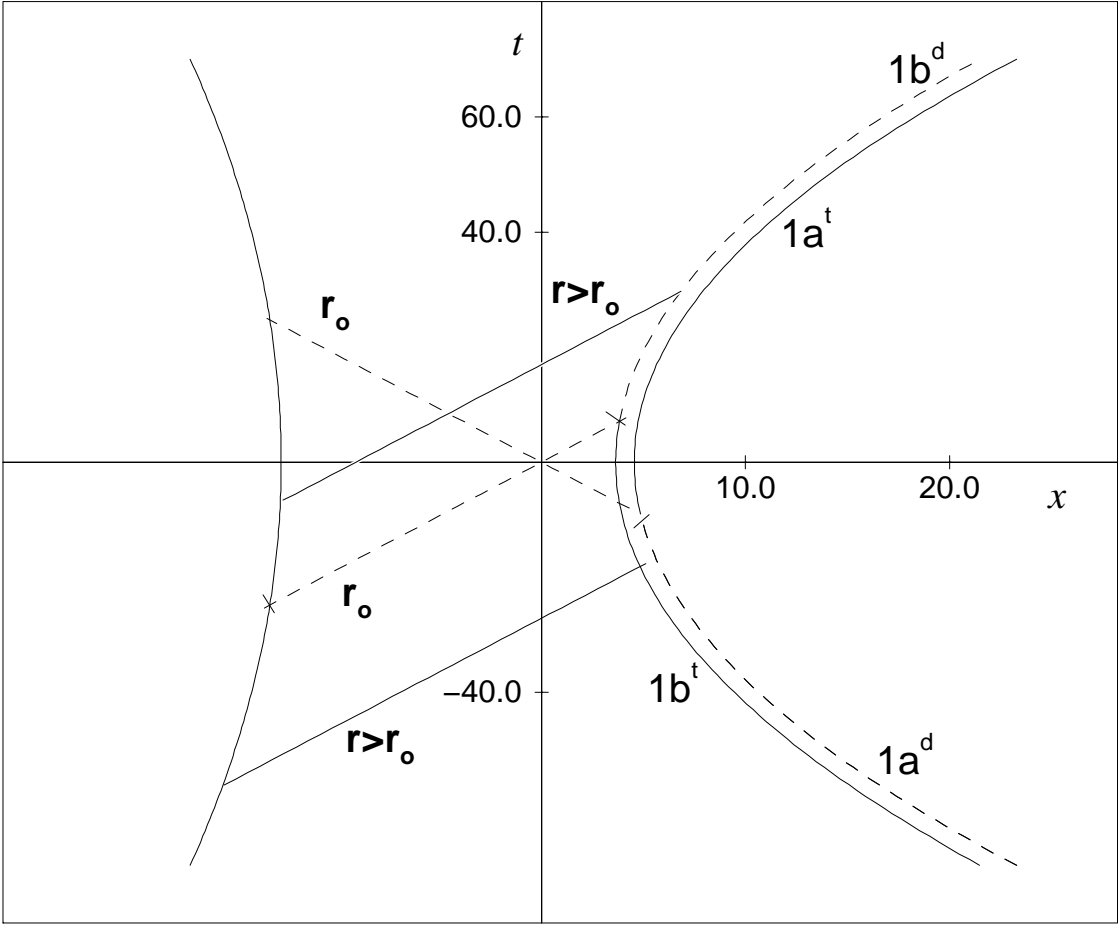
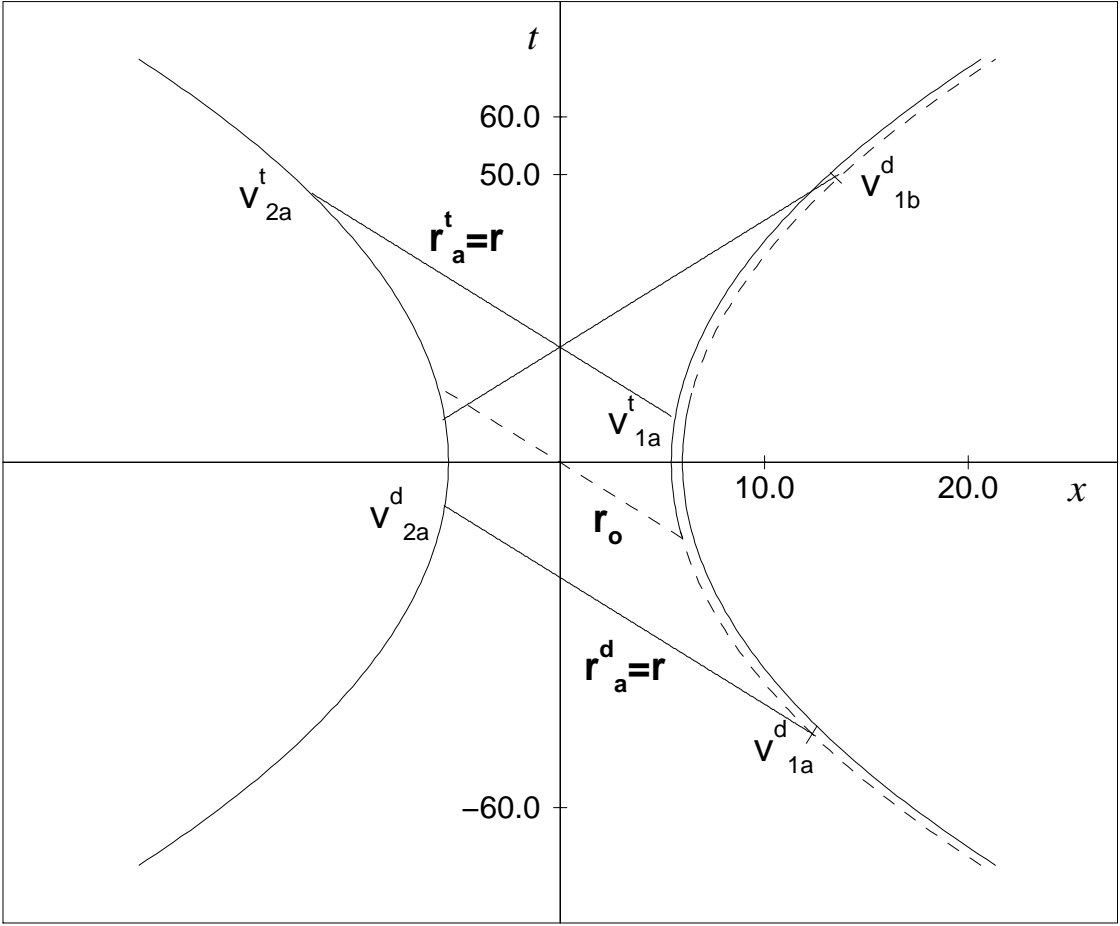
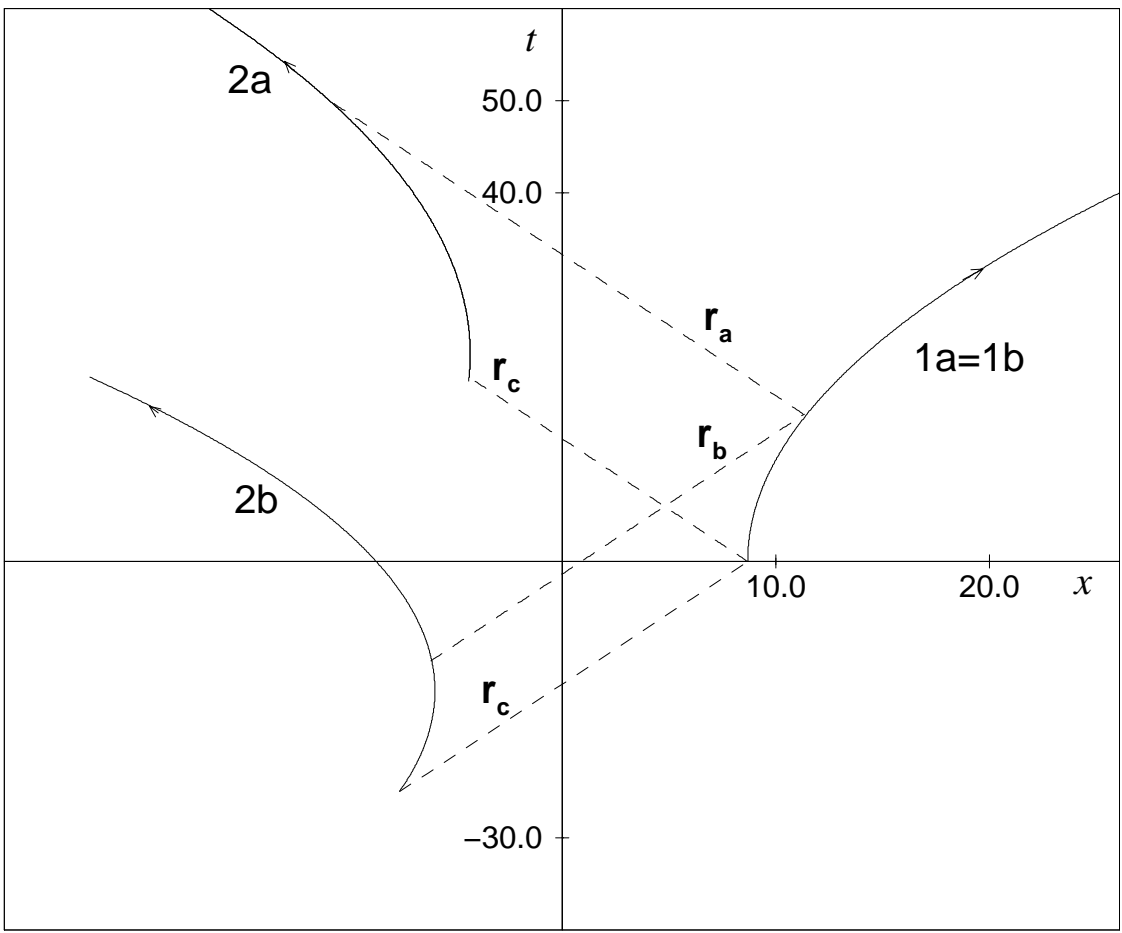
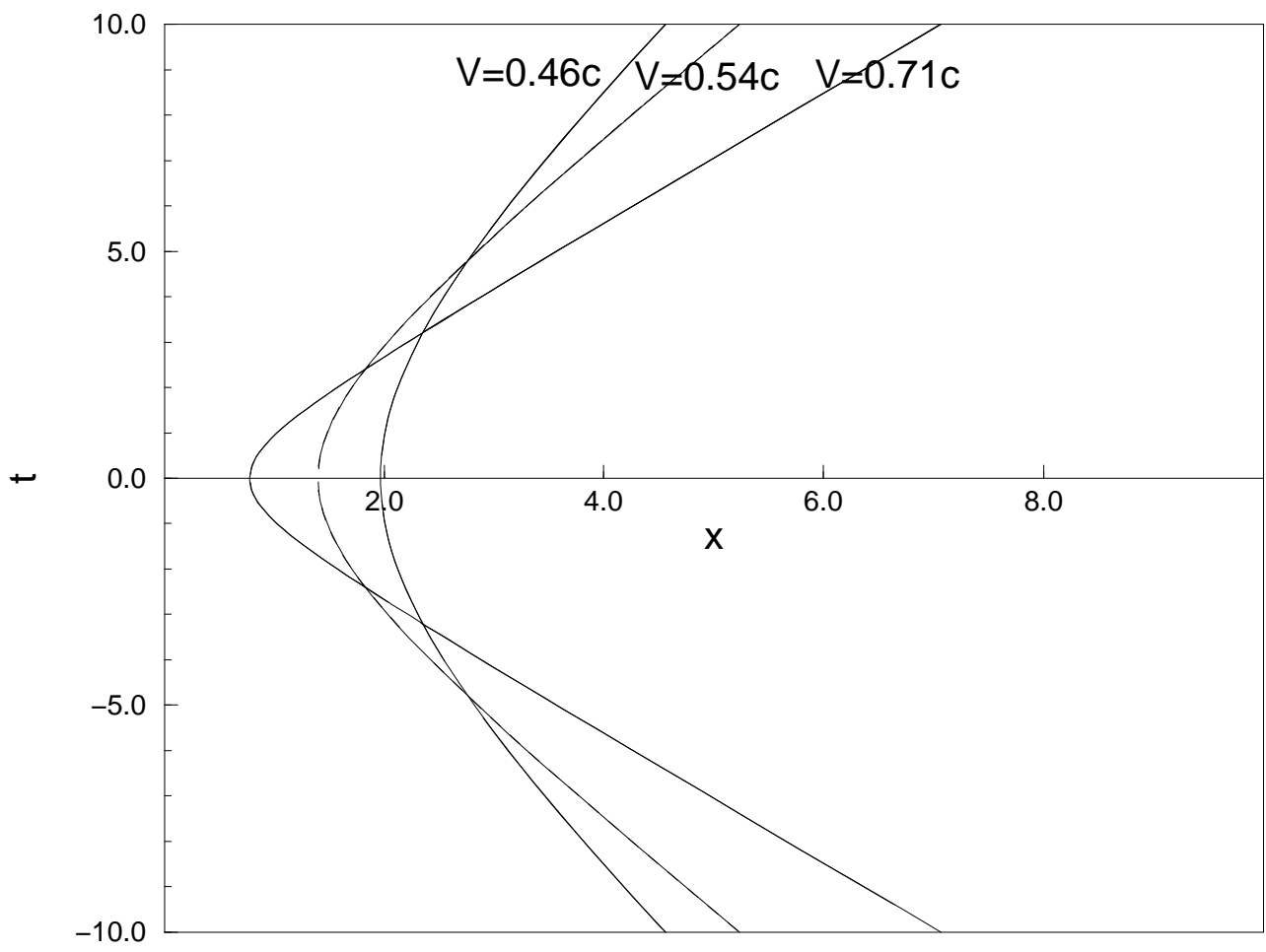
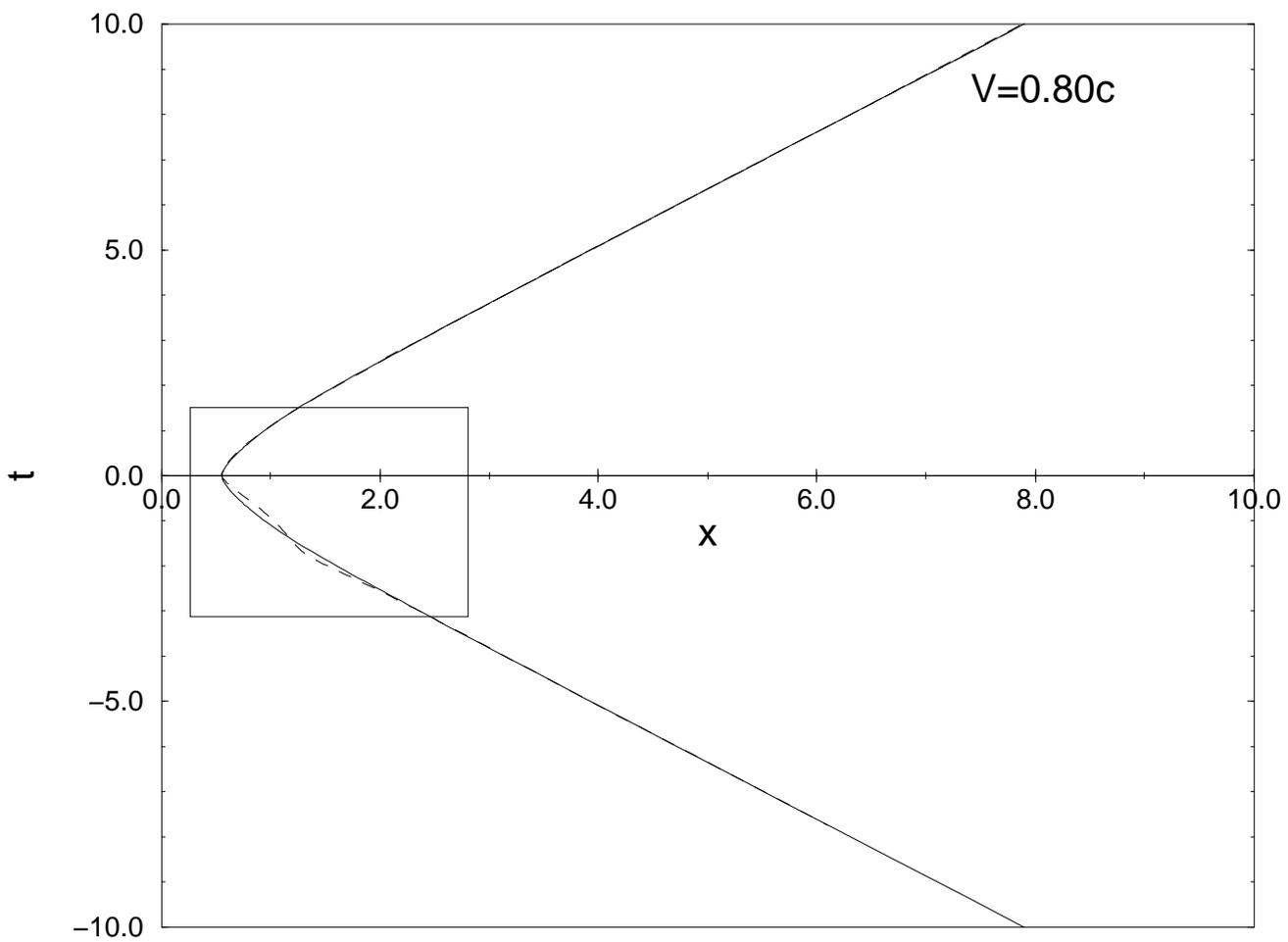
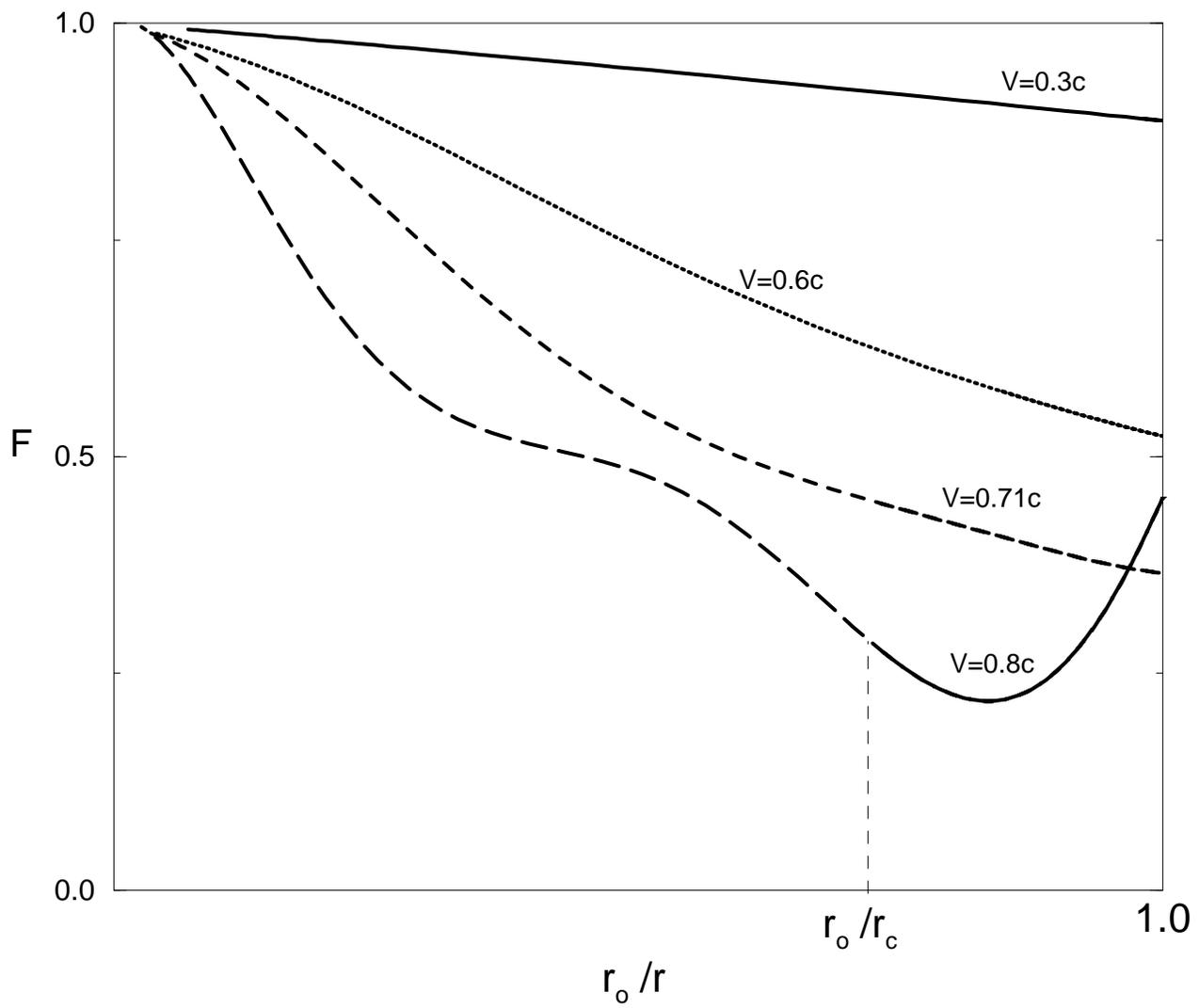
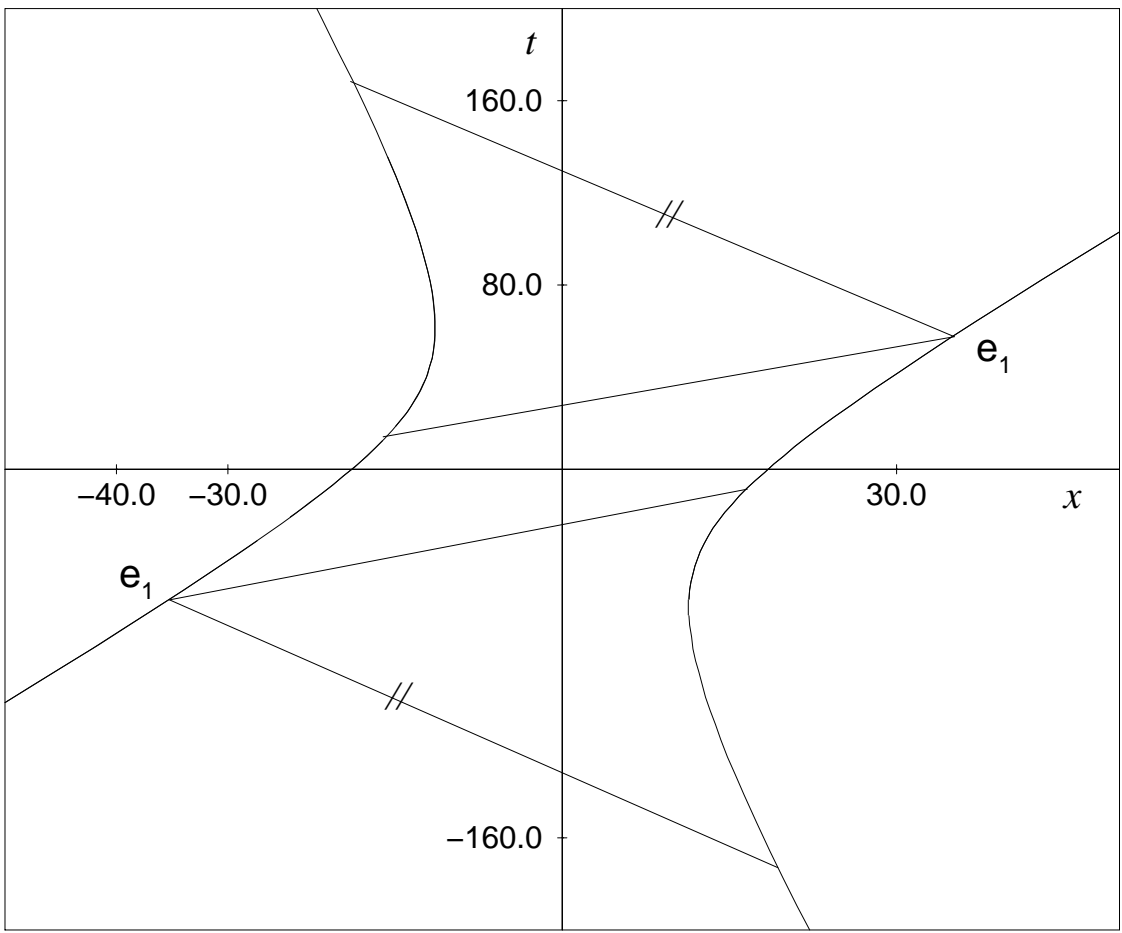
\end{document}